\documentclass[a4paper,fleqn]{cas-sc}

\usepackage[numbers]{natbib}
\usepackage{lipsum} 
\usepackage{multirow}
\usepackage{amsmath,amsfonts,esint}
\usepackage{algorithmic}
\usepackage{array}
\usepackage[caption=false,font=scriptsize,labelfont=sf,textfont=sf]{subfig}
\usepackage{textcomp}
\usepackage{stfloats}
\usepackage{url}
\usepackage{verbatim}
\usepackage{cite}
\usepackage{enumitem}
\usepackage{graphicx,epstopdf,epsfig}
\graphicspath{{figs/}}
\usepackage{hyperref}
\usepackage{cuted}
\usepackage{longtable}
\def\tsc#1{\csdef{#1}{\textsc{\lowercase{#1}}\xspace}}
\tsc{WGM}
\tsc{QE}

\usepackage{xcolor}
\usepackage[linesnumbered,ruled,vlined]{algorithm2e}
\SetCommentSty{mycommfont}
\SetKwInput{KwInput}{Input}     
\SetKwInput{KwOutput}{Output}   
\SetKwInput{KwPar}{Parameter}	

\begin{document}
\let\WriteBookmarks\relax
\def\floatpagepagefraction{1}
\def\textpagefraction{.001}

\shorttitle{} 

\shortauthors{Q. Fu, Z. Qin}

\title [mode = title]{
	Emergence of robust looming selectivity via coordinated inhibitory neural computations
}


%

\author{Qinbing Fu}[orcid=0000-0002-5726-6956]
\fnmark[1]
\cormark[1]
\ead{qifu@gzhu.edu.cn}

\author{Ziyan Qin}[orcid=0000-0001-9215-9331]
\fnmark[1]

\affiliation[1]{
	organization={School of Mathematics and Information Science, Guangzhou University},
	city={Guangzhou},
	postcode={510006}, 
	country={China}
}
\affiliation[2]{
	organization={School of Mathematics and Systems Science, Guangdong Polytechnic Normal University},
	city={Guangzhou},
	postcode={510665}, 
	country={China}
}
\fntext[1]{The authors share first authorship.}

\begin{abstract}
	Inhibitory signal processing is inherent and critical in biological systems, preventing neurons from being easily activated and constraining their activity within an optimal range. 
	Various types of inhibition play a significant role in visual neural circuits, preserving "selectivity" in motion perception. 
	In the locust's lobula giant movement detector (LGMD) neural pathways, four categories of inhibition, i.e., global inhibition, self-inhibition, lateral inhibition, and feed-forward inhibition, have been functionally explored in the context of looming perception, i.e., the neural computation underlying the perception of approaching objects that is crucial for animal survival. However, their combined influence on shaping selectivity to looming motion remains unclear. 
	Driven by recent physiological advancements, this paper offers new insights into the roles of these inhibitory mechanisms at multiple levels and scales in simulations, refining the specific selectivity for responding only to objects approaching the eyes while remaining unresponsive to other forms of movement. 
	Within a feed-forward, multi-layer neural network framework, global inhibition, lateral inhibition, self-inhibition, and feed-forward inhibition are integrated. 
	Global inhibition acts as an immediate feedback mechanism, normalising light intensities delivered by ommatidia, particularly addressing low-contrast looming. 
	Self-inhibition, modelled numerically for the first time, suppresses translational motion. 
	Consistent with previous studies, lateral inhibition is formed by delayed local excitation spreading across a larger area. 
	Notably, self-inhibition and lateral inhibition are sequential in time and are combined through feed-forward inhibition, which indicates the angular size subtended by moving objects. 
	Together, these inhibitory processes attenuate motion-induced excitation at multiple levels and scales. 
	While earlier computational models highlighted the effects of lateral inhibition and feedforward inhibition, this research additionally suggests that (1) self-inhibition may act earlier than lateral inhibition to rapidly reduce excitation in situ, thereby suppressing translational motion, and (2) global inhibition can modulate excitation on a finer scale, enhancing selectivity in higher contrast range. 
	To validate and further explore the proposed inhibition model, we conducted systematic experiments, including ablation studies against a range of visual challenges, and extensively compared our model with the state of the art. 
	The proposed model demonstrated improved selectivity, responding exclusively to approaching stimuli. 
	Moreover, the model operates effectively and robustly on coherence-incoherence stimuli, aligning with recent physiological observations.
\end{abstract}



\begin{keywords}
	biological system model \sep lateral inhibition \sep feed-forward inhibition \sep global inhibition \sep self-inhibition \sep looming selectivity
\end{keywords}

\maketitle

%


\section{Introduction}
\label{Sec: introduction}
The ability to sense and respond to objects on a collision course represents a fundamental survival strategy across the animal kingdom, where visual cues often provide the most reliable source of information. Such stimuli are encoded in the brain through a dynamic balance of excitatory and inhibitory synapses, with rapid dendritic processing that enables the discrimination of complex spatiotemporal patterns.

\begin{figure}[t]
	\centering
	\includegraphics[width=0.4\linewidth,keepaspectratio]{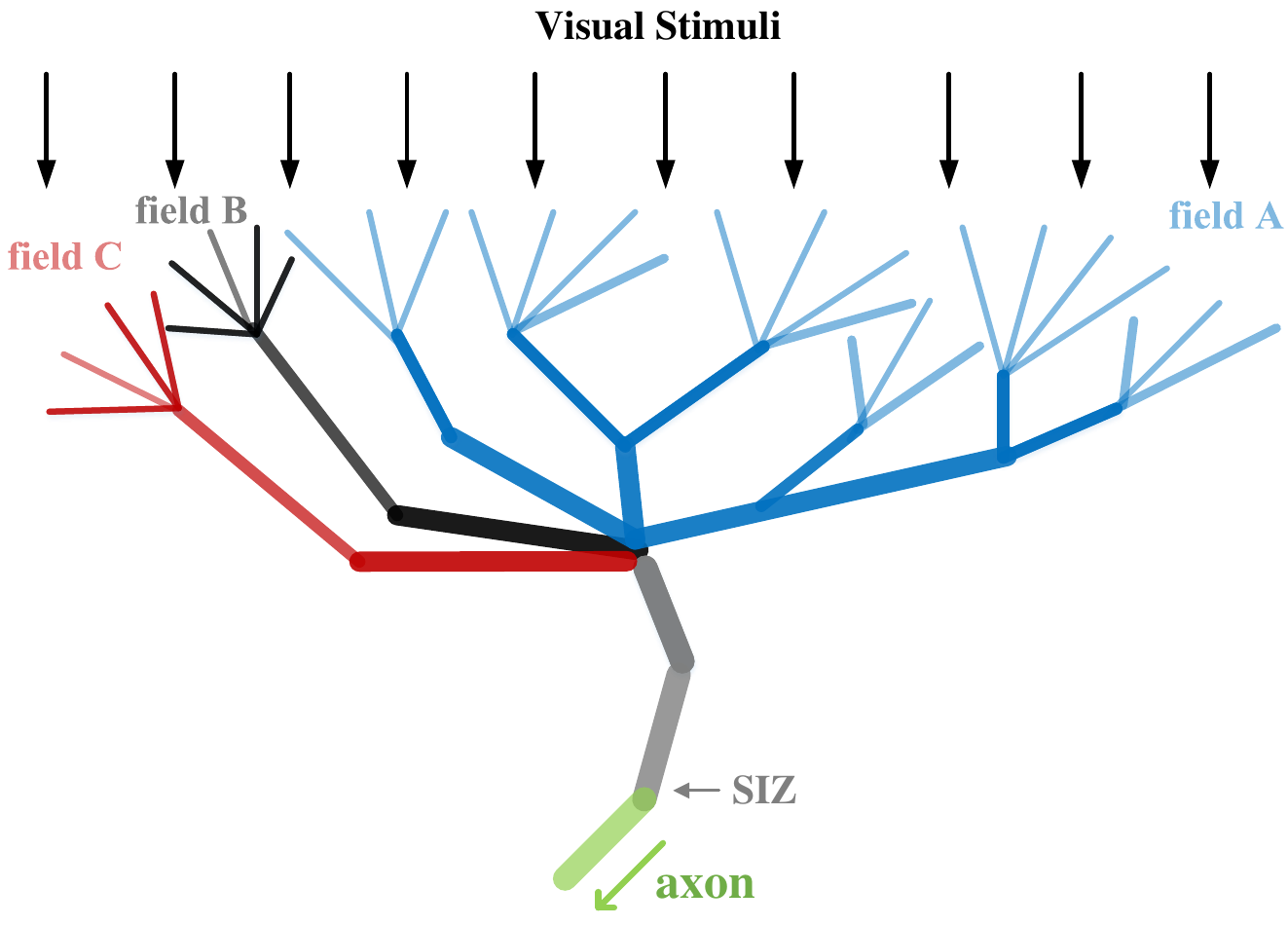}
	\caption{
		\small
		The schematic morphology of dendrites, without metrics, includes three presynaptic fields synapsing to the LGMD. 
		Field A receives visual movement-induced excitations, while fields B and C receive inhibitory signals. 
		The spike initiation zone (SIZ) indicates where the LGMD generates and delivers spikes along the axon.
	}
	\label{Fig: fields morphology}
\end{figure}

A striking example is found in locusts, where the lobula giant movement detector (LGMD) neuron transforms approaching visual signals into robust escape responses, thereby preventing impending collisions \citep{LGMD-1974,Rind1998(local-circuit-locust),LGMD1-1999(Rind-seeing-collision)}.
Intriguingly, the LGMD responds most strongly to the spatiotemporal patterns of objects approaching the eyes, while it is only briefly activated by objects moving across or away from the eyes \citep{LGMD1-1996(Rind-intracellular-neurons)}. 
This capability is termed "\textit{looming selectivity}". 
The mechanisms by which such selectivity has been sculpted and refined through evolutionary development in biological neural circuits remain largely unknown, yet they draw considerable attention from biologists. 
Notably, this attribute is also prominent in flies (e.g., the fruit fly \textit{Drosophila}), although locusts and flies utilize different neural substrates for looming perception \citep{Borst-Review-2019-Fly,Borst-review-2023,Rind-2024-review}.

In neural systems, inhibitory signal processing is intrinsic and plays a critical role in preventing neurons from being easily activated. 
Researchers have gradually shifted their attention to the role of inhibition in the looming perception visual pathways of both locusts and flies, unveiling that it shapes the specific selectivity to objects moving in depth with increasing angular size subtending the insects' compound eyes, known as the receptive field \citep{Mauss-Opposing-Motions-Fly,LPLC-Nature-2017,Wang-LGMD-FFI,Zhu-LGMD-lateral-excitation}.

Currently, there are two main categories of modelling work based on the locust's LGMD. 
The first is the well-established $\eta$-function \citep{Gabbiani2002(multiplicative-computation-LGMD)}, which maps the non-linear relationships between physical stimulus attributes and neuronal activity. 
These attributes include the angular velocity of external stimuli and the angular size subtending the receptive field. 
The second category employs hierarchical neural networks based on the anatomically stratified structure of the optic lobe, including the retina, lamina, medulla, and lobula neuropils, as reviewed in \citep{Fu-ALife-review,Fu-ON/OFF-2023}. 
Although there are ongoing debates on how visual movements are neurally encoded, the importance of inhibition is acknowledged and emphasized in both methodologies.

In the first category, excitatory signals accumulate from the dendrite field A (Fig. \ref{Fig: fields morphology}) to activate the LGMD when an object is approaching with an angular velocity ($\Theta(t)'$), which is the positive derivative of angular size ($\Theta(t)$). 
On the other hand, inhibitory signals, which represent the angular size, are received from the smaller dendrite fields B and C (Fig. \ref{Fig: fields morphology}), and interact non-linearly with the excitatory signals over time \citep{Gabbiani2004(invariance-LGMD),Wang-LGMD-FFI,Contrast-polarity-mapping-2022}. 
This inhibition is referred to as feed-forward inhibition. 
Initially, the computational role of feed-forward inhibition was to directly suppress neuronal activity once the angular size exceeded a threshold size. 
However, recent studies suggested that feed-forward inhibition is mediated by dorsal uncrossed bundle (DUB) neurons within the medulla, which fine-tune the LGMD's firing rate \citep{Wang-LGMD-FFI}. 
To describe the relationship between feed-forward inhibition and feed-forward excitation, the $\eta$-function operates as $\eta(t+\epsilon)=\frac{d \Theta(t)}{d t} \cdot \text{exp}\big(-\alpha\Theta(t)\big)$. 
However, this mathematical function cannot fully explain how the physical attributes, i.e., angular velocity indicated by feed-forward excitation, and angular size indicated by feed-forward inhibition, are decoded by neuropils in the optic lobe. 
Moreover, the $\eta$-function accurately predicts approaching objects at a constant speed \citep{Gabbiani-2023}, but similar phenomenological models are limited to this scenario. The LGMD also shows other notable response preferences, such as transient responses to receding and translating stimuli, and inactivity to wide-field grating or whole-field luminance changes.

Another approach to modelling follows multi-layered neural network structures \citep{Fu-ALife-review,Fu-ON/OFF-2023}. 
Rind and Bramwell proposed a classic framework that encodes external stimuli upon photoreceptors through multiple layers, including excitation, inhibition, and summation layers \citep{LGMD1-1996(Rind-neural-network)}. 
Following this seminal work, numerous computational models mimicking the optic lobe processing of locusts have been proposed. 
These network structures have become increasingly elaborate, with some models capable of handling dynamic and complex collision-detection tasks in driving scenes \citep{Fu-2020-Access,LGMD-car-2017(bionic-vehicle-collision)} and unmanned aerial vehicle scenarios \citep{jiannan-AIAI-2019,jiannan-TNNLS-2021}. 
A common feature of these neural network models is the computational modelling of feed-forward inhibition and lateral inhibition, which cooperate to curtail motion-induced excitation and shape the LGMD's selectivity. 
There are debates on how best to build a model that reproduces physiological responses, but there is consensus on the significance of inhibition in tuning specific looming selectivity. 
Compared to feed-forward inhibition, which can directly shut down the LGMD's activity, lateral inhibirion interacts more gently with local excitation that spreads out in space and time. 
The seminal work by Rind and Bramwell effectively mimicked such interactions in the LGMD's dendrite field A during looming, receding, and translating object movements \citep{LGMD1-1996(Rind-neural-network)}. 
On the other hand, feed-forward inhibition was identified as affecting two smaller, separate dendritic areas of the LGMD, curtailing feed-forward excitation on a global scale due to its wide receptive field \citep{Wang-LGMD-FFI}.

Other types of inhibition have also been explored for their contributions to the specific selectivity of the LGMD, such as global inhibition and self-inhibition. 
Global inhibition has been proposed to normalise excitatory input signals at an early stage in the locust's visual pathways \citep{Zhu-LGMD-lateral-excitation}. 
This type of signal normalisation is a canonical neural computation found in many sensory systems, including olfactory, auditory, and visual systems \citep{Carandini-normalization-in-neural-computation}. 
In the fruit fly \textit{Drosophila}, global inhibition functions similarly by providing instantaneous feedback normalisation, which significantly reduces response variance to dynamic and complex scenes \citep{Drews-dynamic-signal-compression}. 
This feature has been integrated into contrast neural computations to enhance the fidelity of motion vision, particularly in extracting low-contrast movement features \citep{Fu-IJCNN-2021,Fu-Array}. 
Additionally, Rind et al. discovered that trans-medullary afferent neurons (TmAs) connect in a triadic fashion, reciprocally influencing each other \citep{LGMDs-2016}. 
This suggests the presence of self-inhibition presynaptic to the LGMD, which has been shown to effectively suppress translating stimuli.

While several inhibition models have been proposed to incorporate combinations of feed-forward inhibition and lateral inhibition in LGMD models, as well as global inhibition in fly visual systems, significant gaps remain in explaining how these four types of inhibition are interconnected across different neuropils and scales to fine-tune neuronal selectivity to looming motion. 
Additionally, self-inhibition has been less explored in computational modelling, yet it holds potential to enhance the neuronal preference for approaching versus translating stimuli. 
Further research is needed to elucidate these interactions comprehensively and their impact on visual processing in biological systems.

To address these gaps, a multi-scale and multi-level modelling strategy offers a useful framework, as it allows the interplay of inhibition to be examined across both spatial and temporal domains. Comparable approaches have provided valuable insights in broader vision science. For instance, in the visual system, multiscale and multilevel sampling—with or without feedback—has been identified as a key mechanism underlying the emergence of end-stopped cells in V1 layer 4, which in turn contributes critically to solving the aperture problem in area MT \citep{Sherbakov2013}. Similarly, multilayered neural models involving the interaction of excitation and inhibition, together with feedback mechanisms, have successfully simulated perceptual phenomena such as contrast–contrast illusions and apparent motion \citep{ZhuJ2023}. 

\begin{figure}[t]
	\centering
	\includegraphics[width=0.9\linewidth,keepaspectratio]{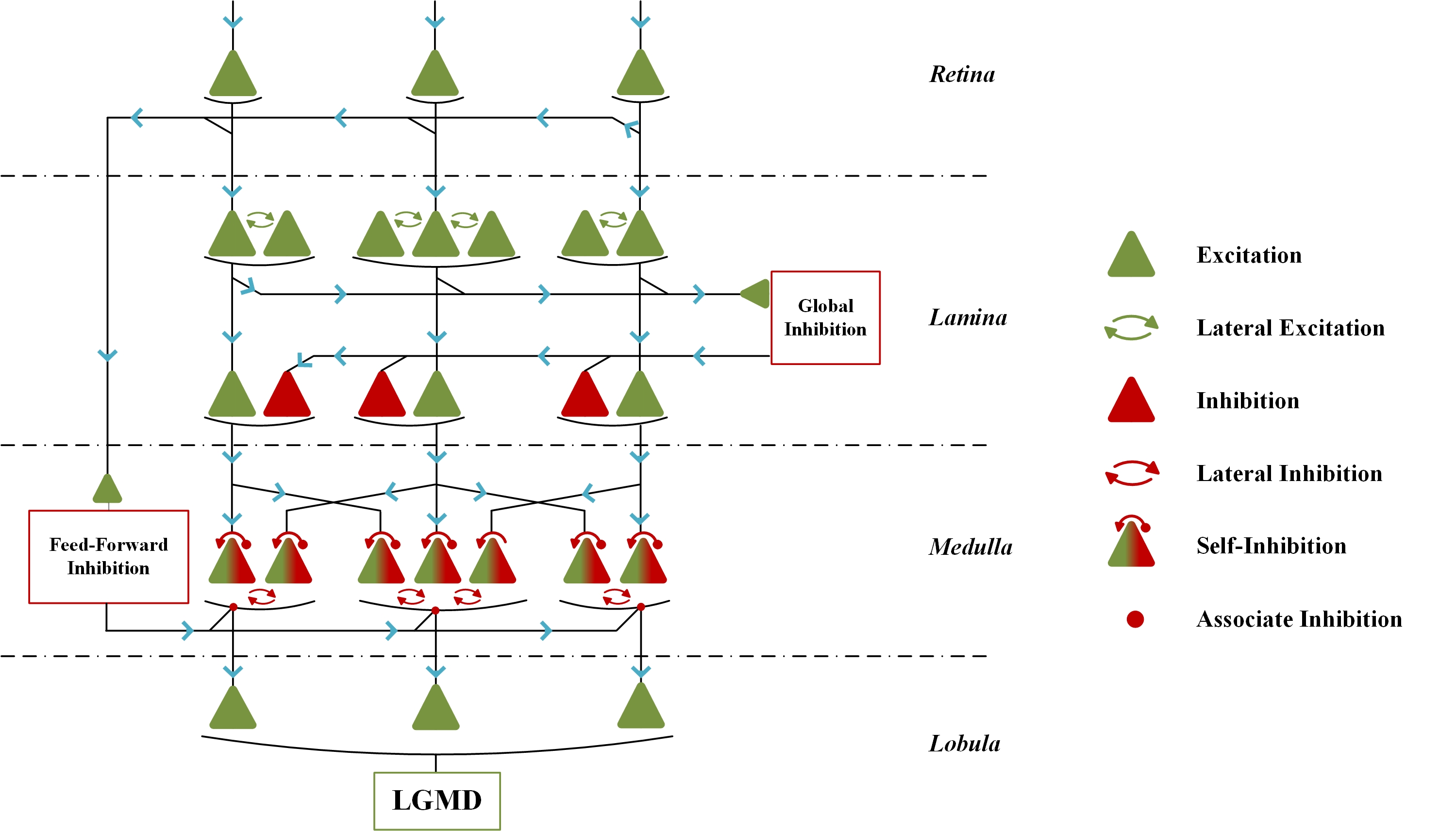}
	\caption{
		\small
		The schematic diagram illustrates multi-level, multi-scale, concomitant inhibitions across four neuropils: retina, lamina, medulla, and lobula, interacting with excitation. 
		Feed-forward  inhibition operates globally, indicating angular size and mediating the effects of self-inhibition and lateral inhibition in the medulla. 
		global inhibition normalizes excitations to a specific range. 
		Self-inhibition curtails local excitation before it spreads to neighbouring fields. 
		Lateral inhibition competes with excitation on a larger scale than self-inhibition. 
		The remaining excitatory signals are eventually integrated by the LGMD cell.
	}
	\label{Fig: four inhibitions}
\end{figure}

In this work, we put forward a neural computational model in which looming perception is mediated by the coordinated action of lateral, feed-forward, global, and self-inhibition (Fig. \ref{Fig: four inhibitions}). 
Fig. \ref{Fig: four inhibitions} illustrates the proposed model in a stratified neural network. 
Compared to related modelling work, the emphasis herein is laid upon the computational roles of four inhibitions. 
Feed-forward inhibition acts across the whole receptive field indicating the angular size, affecting self-inhibition and lateral inhibition at the third neuropil of optic lobe, i.e., medulla. 
Global inhibition normalizes lateral excitation relayed to lamina, at the relatively larger scale than self-inhibition and lateral inhibition. 
To align with the physiological research of lateral inhibition and self-inhibition \citep{LGMDs-2016}, we propose coordinated neural computations of lateral inhibition and self-inhibition curtailing local excitation. 
Taken together, this research provides the following new perspectives:
\begin{itemize}[leftmargin=*]
	\item The four inhibitory mechanisms can be effectively integrated within a neural network framework to collectively shape specific selectivity towards approaching stimuli.
	\item This inhibition model operates through sequential interactions of lateral inhibition and self-inhibition with excitation. 
	Lateral inhibition and self-inhibition operate in a complementary fashion, counterbalancing each other in a way that is further regulated by feed-forward inhibition.
	Initially, self-inhibition suppresses local excitation locally, but its impact diminishes as the stimulus size (angular size) increases. 
	Consequently, self-inhibition plays a crucial role in processing stimuli with nearly constant angular size. 
	On the other hand, lateral inhibition results from delayed excitation spreading spatially, making its influence more pronounced with increasing AS. 
	Thus, lateral inhibition effectively distinguishes approaching from receding stimuli. 
	This modelling approach aligns with Rind's perspective in \citep{LGMDs-2016}.
	\item Global inhibition plays a unified role in dynamically suppressing light intensities across its neighbouring fields instantaneously, thereby stabilizing low-contrast looming perception.
	\item Coherent and incoherent motion were comprehensively investigated through comparative and ablation studies. 
	The proposed model effectively detects inconsistent approaching stimuli, outperforming the state of the art, while not responding to other types of movements.
\end{itemize}

The rest of this paper is organised into several sections. 
Section \ref{Sec: neuroscience} reviews related biological research. 
Section \ref{Sec: model} elucidates the proposed method with formulation, algorithm, setting of parameters and experiments. 
Section \ref{Sec: functionality} demonstrates basic functionality of the proposed model against a variety of visual stimuli ranging from synthetic to real-world scenes. 
Section \ref{Sec: coherence} investigates and compares the proposed method with state-of-the-art models against coherence-incoherence stimuli. 
Section \ref{Sec: ablation} provides insights into each kind of inhibition through ablation studies. 
Section \ref{Sec: discussion} gives further discussions. 
Section \ref{Sec: conclusion} concludes this research.

\section{Biological System Research}
\label{Sec: neuroscience}
This section concisely reviews our background study based on biological system research upon inhibition in looming perception. 
Inhibition is normally generated by motion-induced excitation that spreads out in space and time, subsequently interacting with succeeding excitation in place. 
Consequently, spatiotemporal inhibition works in close concert with excitation to shape the specific firing patterns of motion-sensitive neurons. 
These neurons are categorized by visual feature selectivity, including looming-responsive neurons \citep{LPLC-Nature-2017, LGMDs-2016}, object detection neurons \citep{Fly2017(object-detecting-neuron)}, and are further ordered by preferred visual feature size \citep{Fly-LCs-2022}.

Specifically, in the locust's visual pathways for looming perception, early studies nearly three decades ago identified the lobula giant movement detector (LGMD)–descending contralateral movement detector (DCMD) pathway, which plays a critical role in collision perception and evasion \citep{DCMD-1992(selective-response-approaching),DCMD-1997(collision-trajectories)}.
A notable characteristic of this neuronal pathway is its strong selectivity for looming objects. 
To explain how this selectivity is sculpted, Rind and Bramwell provided insights into the role of inhibition in a four-layered neural network resembling the stratified optic lobe \citep{LGMD1-1996(Rind-neural-network)}. 
They highlighted that lateral inhibition shapes the velocity tuning of the LGMD network for critical image cues, such as expanding object edges, and is crucial for the selective response to approaching stimuli versus translational or receding stimuli. 
During object motion, lateral inhibition is formed by delayed excitation spreading laterally in space, which curtails fast-building excitation in situ. 
A critical race occurs between excitation and lateral inhibition, where the excitation must prevail to propagate through the network. 
In contrast, feed-forward inhibition is most significant near the end of an approach or at the initiation of recession, directly suppressing the neuron when the angular size is large in both cases. 
Eliminating the feed-forward inhibition would prolong the LGMD's activity during both approach and recession.

Recent progress has expanded our understanding of inhibition in looming perception. 
Zhu and colleagues identified the presence of global inhibition upstream of lateral inhibition, which can normalise the strength of excitatory inputs to the LGMD \citep{Zhu-LGMD-lateral-excitation}. 
This functionality is similar to that documented in fly visual systems \citep{Drews-dynamic-signal-compression}. 
Moreover, Rind et al. reconstructed the presynaptic network to the LGMD, revealing that TmA neurons connect per facet per LGMD and laterally with other TmA neurons \citep{LGMDs-2016}. 
This indicates that TmA neurons can immediately inhibit themselves and their neighbours, which is the origin of self-inhibition. 
These reciprocal inhibitory effects can enhance the LGMD's selectivity for approaching objects over passing ones. 
However, self-inhibition loses its efficacy when the angular size becomes large, i.e., when objects get closer to the receptive field. 
At this moment, lateral inhibition becomes more effective.


\section{Methods}
\label{Sec: model}

This section presents a systematic formulation with emphasis on the associative inhibitions involving feed-forward, global, self, and lateral inhibition in the context of looming perception (Fig. \ref{Fig: signal processing diagram}).  For clarity and brevity, the four types of inhibition will be abbreviated as FFI, GI, SI, and LI, respectively, throughout Sections  \ref{Sec: model} to \ref{Sec: ablation}.
Parametric and experimental settings are also elaborated.

\subsection{Framework of the Inhibition Model}
\begin{figure}[t]
	\vspace{-10pt}
	\centering
	\includegraphics[width=0.5\linewidth,keepaspectratio]{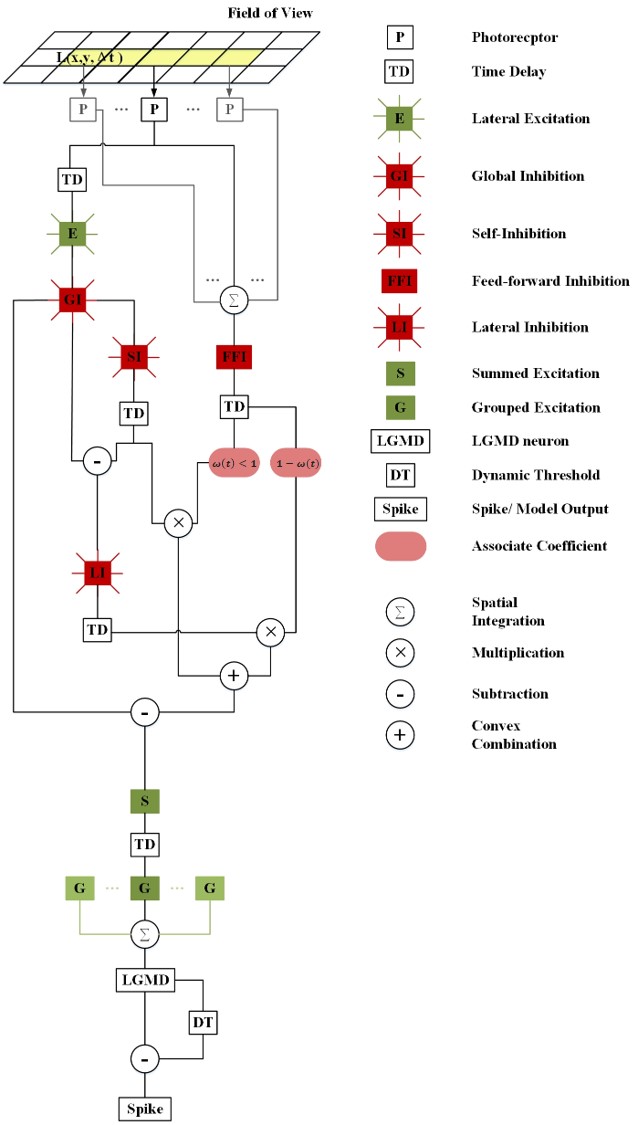}
	\caption{
		\small
		This schematic illustrates the interactions between different types of inhibition and excitation, shaping the response of the LGMD cell in visual signal processing. 
		FFI acts globally to encode angular size and governs the time-dependent trade-off between LI and SI. 
		GI normalizes lateral excitation upstream of LI and SI. 
		SI initially reduces local excitation at the smallest scale. 
		LI subsequently reduces remaining excitation. 
		LGMD cell sieves and absorbs grouped excitations. 
		Dynamic threshold mechanism is applied during spike initiation.
	}
	\label{Fig: signal processing diagram}
\end{figure}

The inhibitions operate across multiple layers mimicking the insect's optic lobe, including retina, lamina, medulla, and lobula neuropils (Fig. \ref{Fig: signal processing diagram}). 
In contrast to typical LGMD-based models \citep{LGMD1-Glayer(feature-enhancement),Hu-2017(Colias-LGMD1)}, where FFI enforces an "all-or-none" response by directly shutting down neurons in response to large, instantaneous luminance changes, the proposed model integrates FFI differently. 
FFI acts globally across the entire receptive field at the largest scale, indicating angular size and temporally modulating LI and SI. 
Inspired by research on flies \citep{Drews-dynamic-signal-compression} and locusts \citep{Zhu-LGMD-lateral-excitation,Wang-LGMD-FFI}, GI functions as instantaneous feedback normalisation, suppressing cell excitation based on the activation of neighbouring regions. 
SI, located in the medulla, rapidly suppresses local excitation but becomes less effective when angular size is large, such as during critical moments like the end of approach or the beginning of recession \citep{LGMDs-2016}. 
LI follows SI, shaping looming selectivity, and its computational role aligns with previous modelling findings \citep{Fu-ALife-review}.

In a departure from previous approaches, LI and SI engage with excitation in a time-dependent, mutually compensatory way, a mechanism supported by FFI.
As angular size increases, LI progressively reduces local excitation. 
This cascade of inhibitions effectively distinguishes approaching objects from translational-like movements (e.g., grating, elongation, translation).

\subsubsection{Encoding external visual stimuli}
The external visual movements are handled via a differential image encoder. 
This mimics the retina layer of the compound eyes of insects, in which each photoreceptor independently collects light signals within the receptive field. 
As a simplified computational model, a matrix of $R$-rows and $C$-columns is used to simulate the arrangement of photoreceptors. 
The input is $L(x,y,t)\in \mathbb{R}^{3}$ where $x$ and $y$ represent abscissa and ordinate of the matrix, and $t$ is the temporal position of the visual streams. 
Mathematically, such a process can also be expressed in a continuous form:
\begin{equation}
	P(x,y,t) =\int \vert \big(\delta(t-\tau) - \delta(t-\Delta t-\tau)\big) \vert L(x, y, \tau)d\tau, \nonumber
\end{equation}
where $P(x,y,t)$ denotes the luminance change captured by the photoreceptor, $\delta$ is the unit impulse function and $\Delta t$ is the time interval between successive frames of input digital signals, $\tau$ is the time constant. 
For digital signals, the time is discrete and the differential encoder can be defined as 
\begin{equation}
	P(x,y,t) = \vert L(x,y,t) - L(x,y,t-1) \vert + \sum_{i=1}^{N_p}a_i P(x,y,t-i),
	\label{Eq:P}
\end{equation}
where $N_p$ stands for the number of persistent frames and $a_i$ is the decay coefficient calculated by
\begin{equation}
	a_i = \left(1 + e^{i}\right)^{-1}. \nonumber
\end{equation}

Following that, a Gaussian blur is applied through spatial filtering as 
\begin{equation}
	\hat{P}(x,y,t) = \iint P(u,v,t)G_{\sigma}(x-u,y-v)dudv, \nonumber
\end{equation}
and discretely as 
\begin{equation}
	\hat{P}(x,y,t) = \sum_{u=-1}^{1}\sum_{v=-1}^{1} P(x-u,y-v,t) \cdot G_{\sigma}(u,v),
	\label{Eq:Blur}
\end{equation}
where $\sigma$ is the standard deviation and the Gaussian kernel can be obtained by
\begin{equation}
	G_{\sigma}(u,v) = \frac{1}{2\pi \sigma^2} \text{exp}\left( -\frac{u^2 + v^2}{2 \sigma^2} \right). \nonumber
\end{equation}

\subsubsection{Feed-forward inhibition indicating angular size}
Over the past two decades, extensive biological research has highlighted the role of FFI in indicating the angular size of moving objects within the receptive field \citep{Gabbiani-2001(LGMD-invariance-angular),Gabbiani-2002(LGMD-multiplicative-computation),Gabbiani2004(invariance-LGMD),Wang-LGMD-FFI,Contrast-polarity-mapping-2022}. 
Unlike earlier models of the locust's LGMD visual pathways, where FFI directly suppresses LGMD neuronal activity, the computational role of FFI in current models is integrated with SI and LI. 
These interactions are typically described by a series of equations:
\begin{equation}
	F(t) = \sum_{x=1}^{R}\sum_{y=1}^{C} P(x,y,t) \cdot (C\cdot R)^{-1},
	\label{Eq:FFI}
\end{equation}
\begin{equation}
	\hat{F}(t) = \alpha F(t) + (1-\alpha) \hat{F}(t-1),\ \alpha = \Delta t / (\tau + \Delta t),
	\label{Eq:FFI-delay}
\end{equation}
where $C$ and $R$ indicate the columns and rows of visual stimuli consistent with the spatial dimensions of neural network layers. 
According to Eq. \ref{Eq:FFI-delay}, the FFI is temporally delayed, which can also be expressed in a differential equation as 
\begin{equation}
	\frac{d \hat{F}(t)}{d t} = \frac{1}{\tau} (F(t) - \hat{F}(t)). \nonumber
\end{equation}
As the FFI increases, the angular size rises monotonically reminiscent of the object getting closer to the observer. 
Subsequently, the FFI is transformed to a time-varying coefficient $\omega$ associated with SI and LI, in a logarithmic manner 
\begin{equation}
	\omega(t) = 1 / \log(\hat{F}(t)),\ s.t. \ \hat{{F}(t)} >= \exp(1).
	\label{Eq:AS}
\end{equation}

\subsubsection{Global normalising inhibition}
GI is a prevalent feature across various sensory systems, including those of visual and olfactory organisms \citep{Carandini-normalization-in-neural-computation}. 
Zhu et al. proposed that GI exists within the presynaptic dendritic area of LGMD's excitatory afferents \citep{Zhu-LGMD-lateral-excitation}. 
Computational studies have shown that GI functions similarly to its physiological role observed in flies \citep{Drews-dynamic-signal-compression}. 
The spatially filtered signals undergo an instantaneous feedback operation as dynamic normalisation
\begin{equation}
	M(x,y,t) = \tanh(\frac{\hat{P}(x,y,t)}{\bar{P}(x,y,t) + \beta}).
	\label{Eq:GI}
\end{equation}
The dynamic formulation of GI with instant, divisive contrast adaptation accounts for robustness in fast changing visual environments. 
\text{$tanh$} operation indicates the hyperbolic tangent function. 
The coefficient $\beta$ determines the baseline sensitivity. 
$\bar{P}(x,y,t)$ can be obtained by Gaussian convolution of surrounding cell responses as 
\begin{equation}
	\bar{P}(x,y,t) = \sum_{u=-5}^{5}\sum_{v=-5}^{5} P(x-u,y-v,t) \cdot G_{\sigma}(u,v),
	\label{Eq:GI-conv}
\end{equation}

\subsubsection{Self-inhibition}
\begin{figure}[t]
	\centering
	\includegraphics[width=0.65\linewidth,keepaspectratio]{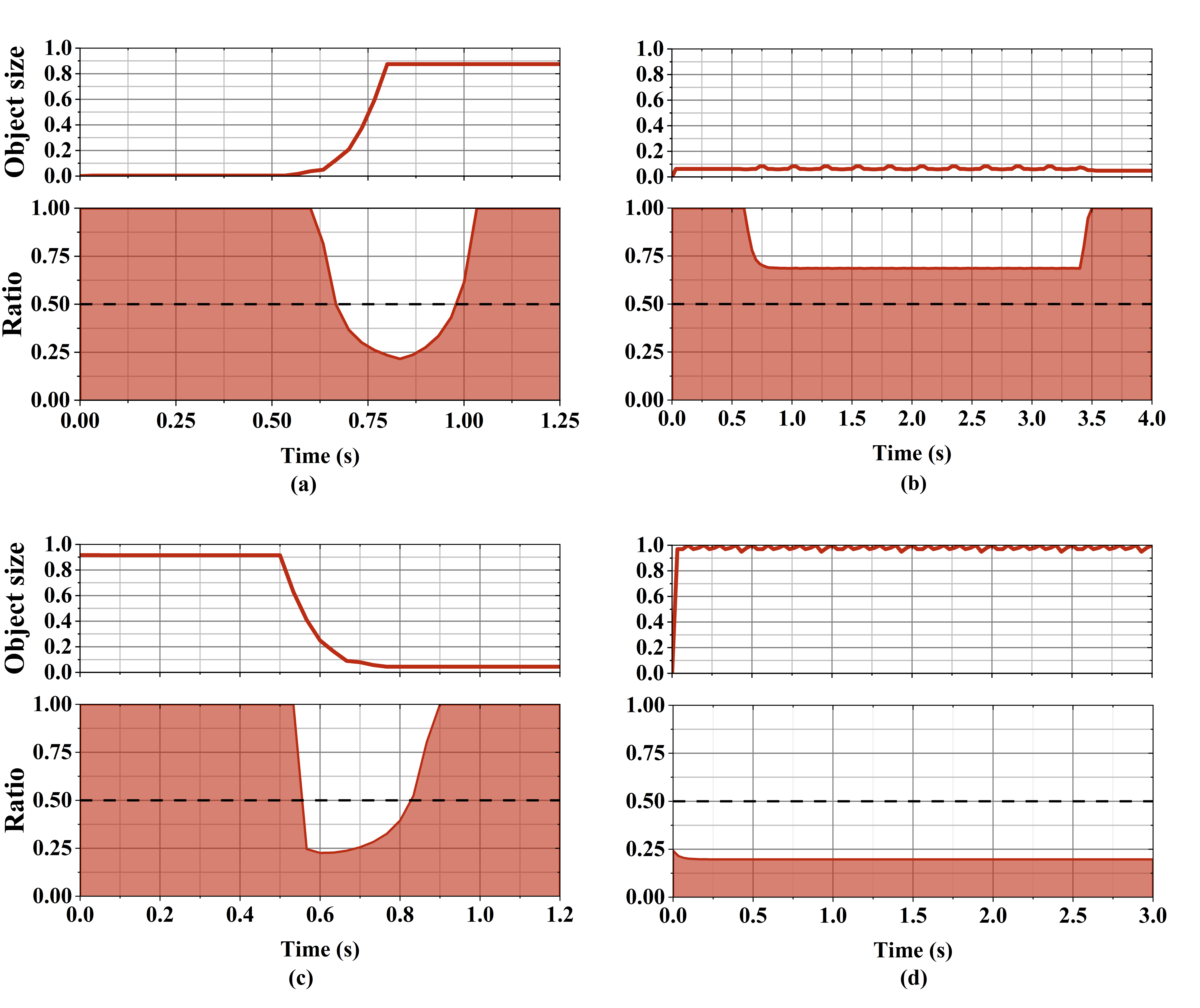}
	\caption{
		\small
		The time-varying coefficient computed by FFI in Eq. \ref{Eq:AS} is associated with LI and SI in a complementary fashion. 
		The red and white shading indicates the respective weights of SI and LI over time during (a) approaching stimuli, (b) translating stimuli, (c) receding stimuli, and (d) grating stimuli. 
		The object size is defined as the ratio of the moving object’s area relative to the entire visual field.
	}
	\label{Fig: SI-LI coeficcient variation}
\end{figure}
Rind and colleagues elucidated the role of TmA neurons by reconstructing the network of input synapses onto LGMD1 and LGMD2, alongside simulating SI and LI using a neural network \citep{LGMDs-2016}. 
SI was observed to play a pivotal role in shaping specific looming selectivity: it effectively reduces neuronal activity at the onset of translation but is overcome by excitation during object approach, contrasting with the delayed and broader action of LI. 
Here we apply a Gaussian convolution at the smallest scale of neighbouring field to obtain the SI. 
The local, normalised excitation is convolved through
\begin{equation}
	SI(x,y,t) = \sum_{u=-1}^{1}\sum_{v=-1}^{1} M(x-u,y-v,t) \cdot G_{\sigma}(u,v),
	\label{Eq:SI-convolve}
\end{equation}
where the kernel conform to Eq. \ref{Eq:Blur}. 
The SI is delayed and at the first stage cuts back the excitation. 
The computations are defined as the following
\begin{equation}
	\hat{SI}(x,y,t) = \alpha SI(x,y,t) + (1-\alpha) \hat{SI}(x,y,t-1),
	\label{Eq:SI-delay}
\end{equation}
where the coefficient conforms to Eq. \ref{Eq:FFI-delay}. 
There is a passing layer that activates the interacted signals as
\begin{equation}
	\hat{M}(x,y,t) = \big[M(x,y,t)-\hat{SI}(x,y,t)\big]^{+},
	\label{Eq:SI-passing}
\end{equation}
where $[x]^{+}$ is described mathematically as
\begin{equation}
	[x]^{+} = \max(x, 0). \nonumber
\end{equation}
Accordingly, the SI acts earlier than the LI to curtail the local excitation.

\subsubsection{Lateral inhibition}
LI is a fundamental mechanism in sensory systems, crucial in the visual pathways for looming perception. 
LI is characterized by the delayed spread of excitation, which curtails newly generated excitation to prioritize looming over recession, translation, and wide-field grating movements \citep{LGMD1-1996(Rind-neural-network),Simmons-1997(LGMD2-neuron-locusts)}. 
LI operates at a larger spatial scale compared to SI, acting later in the processing sequence. 
Importantly, LI suppresses activity in its surrounding area rather than SI. 
As a result, the remaining excitation after SI-sieving undergoes another convolution process as 
\begin{equation}
	\begin{aligned}
		LI(x,y,t) &= \sum_{u=-2}^{2}\sum_{v=-2}^{2} \hat{M}(x-u,y-v,t) \cdot \hat{G}_{\sigma}(u,v),\\
		&s.t. \ \hat{G}_{\sigma}(0,0)=0.\\
	\end{aligned}
	\label{Eq:LI-convolve}
\end{equation}
The LI then is delayed for a same time window as 
\begin{equation}
	\hat{LI}(x,y,t) = \alpha LI(x,y,t) + (1-\alpha) \hat{LI}(x,y,t-1).
	\label{Eq:LI-delay}
\end{equation}

\subsubsection{Concomitant inhibitions}
As presented above, SI and LI are associated with FFI to interact with excitation. 
Mathematically, 
\begin{equation}
	\begin{aligned}
		&S(x,y,t) =\\ &\big[M(x,y,t)-\omega(t) \cdot \hat{SI}(x,y,t)-(1-\omega(t)) \cdot \hat{LI}(x,y,t)\big]^{+},\\
	\end{aligned}
	\label{Eq:LI-passing}
\end{equation}
then delayed by 
\begin{equation}
	\hat{S}(x,y,t) = \alpha S(x,y,t) + (1-\alpha) \hat{S}(x,y,t-1).
	\label{Eq:S-delay}
\end{equation}

To examine how inhibition varies with different types of motion, we considered four categories of visual stimuli: approach, recession, translation, and grating. In this context, approach refers to an object expanding in size on the retina, indicating motion toward the observer, whereas recession denotes the opposite, with the object shrinking in size and thus moving away. Translation describes lateral displacement across the visual field, independent of approach or recession, while grating is a specific form of translation in which a bar moves laterally with varying temporal and spatial frequencies. 

Fig.~\ref{Fig: SI-LI coeficcient variation} exemplifies how the coefficients of SI and LI change under these different motion conditions. The analysis shows that the weight of LI increases as the angular size grows, while the weight of SI correspondingly decreases. For translation-like stimuli, SI remains relatively stable. This pattern of modulation is consistent with the findings reported by Rind and colleagues \citep{LGMDs-2016}.
It is important to note that while the FFI-regulated combination of SI and LI represents the optimal setting in our proposed model, we do not rule out the possibility that SI and LI could occur successively without combination, or that they may interact in other ways.

\subsubsection{Integrating feed-forward excitation}
After competition between local excitations and combined inhibitions, the remaining feed-forward excitation (FFE) is further grouped to reduce isolated noise in cluttered backgrounds. 
This is implemented with a passing coefficient matrix $[S_e]$, determined by a convolution with an equally weighted kernel, as the following
\begin{equation}
	S_e(x,y,t) = \sum_{i=-1}^{1}\sum_{j=-1}^{1} \hat{S}(x+i,y+j,t)\cdot W_s(i,j),
	\label{Eq:S-conv}
\end{equation}
\begin{equation}
	W_s = \frac{1}{9}\ \times \left[
	\begin{matrix}
		1 & 1 & 1\\
		1 & 1 & 1\\
		1 & 1 & 1
	\end{matrix}
	\right],
	\label{Eq:S-matrix}
\end{equation}
\begin{equation}
	G(x,y,t) = \hat{S}(x,y,t)\cdot S_e(x,y,t)\cdot \rho(t)^{-1},
	\label{Eq:G}
\end{equation}
\begin{equation}
	\rho(t) = \text{max}([S_e]_t)\cdot C_{\rho}^{-1} + \Delta_C.
	\label{Eq:G-scale}
\end{equation}
\begin{equation}
	\hat{G}(x,y,t)=\left\{\begin{array}{ll}
		G(x,y,t) & \text { if } G(x, y,t) \ge T_{g}   \\
		0 & \text { otherwise }
	\end{array}\right.
	\label{Eq:G-threshold}
\end{equation}

In the lobula area, the LGMD cell integrates all pre-synaptic FFE so as to generate the membrane potential as the following
\begin{equation}
	k(t) =\int_{1}^{R}\int_{1}^{C} \hat{G}(x,y,t)dxdy \nonumber
\end{equation}
and discretely, 
\begin{equation}
	k(t) = \sum_{x=1}^{R}\sum_{y=1}^{C} \hat{G}(x,y,t),\ K(t) = \left(1 + e^{-k(t)\cdot (C \cdot R \cdot \gamma)^{-1}}\right)^{-1},
	\label{Eq:activation}
\end{equation}

Differently to previous neural network models whereby a fixed potential threshold participates in eliciting spikes, the proposed model adopts a temporally dynamic spiking threshold with regard to a latest model \citep{Qin-Fu-DNF-2024}. 
\begin{equation}
	K_{thre}(t) = \frac{\sum_{t-N_{dt}}^{t}K(t)}{N_{dt}}
	\label{Eq:dynamic threshold}
\end{equation}

Finally, a spike would be generated if the current membrane potential oversteps the dynamic threshold:
\begin{equation}
	\text{Spike}(t) = 
	\begin{cases}
		1, & {\text{if}} \ K(t) > K_{thre}(t),\\
		{0,} & {\text{otherwise}}.
	\end{cases}
	\label{Eq:spike}
\end{equation}

\begin{algorithm}[t]
	\DontPrintSemicolon
	\KwInput{ $L(x,y,t) \in \mathbb{R}^{3}$ }
	\KwOutput{$\text{Spike}(t)$}
	\KwPar{Table~\ref{Tab: parameters}}
	\For{t = 1:T}{
		\tcp{Input}
		Receive Input $L(x,y,t)$
		
		\tcp{Photoreceptor response} 
		Calculate $P(x,y,t)$ via Eq. (\ref{Eq:P})
		
		\tcp{Gaussian blur}		
		Calculate $\hat{P}(x,y,t)$ via Eq. (\ref{Eq:Blur})
		
		\tcp{FFI tuning time-varying angular size indicator}
		Calculate $\hat{F}(t)$ to obtain $\omega(t)$ via Eq. (\ref{Eq:FFI},\ref{Eq:FFI-delay},\ref{Eq:AS})
		
		\tcp{Global inhibition}
		Calculate $M(x,y,t)$ via Eq. (\ref{Eq:GI},\ref{Eq:GI-conv})
		
		\tcp{Self inhibition}
		Calculate $SI(x,y,t)$ via Eq. (\ref{Eq:SI-convolve},\ref{Eq:SI-delay})
		
		\tcp{Lateral inhibition}
		Calculate $LI(x,y,t)$ via Eq. (\ref{Eq:SI-passing},\ref{Eq:LI-convolve},\ref{Eq:LI-delay})
		
		\tcp{Coordination of SI and LI lessening excitation}
		Calculate $S(x,y,t)$ via Eq. (\ref{Eq:LI-passing},\ref{Eq:S-delay})
		
		\tcp{Grouped remaining excitation}
		Calculate $G(x,y,t)$ via Eq. (\ref{Eq:S-conv},\ref{Eq:S-matrix},\ref{Eq:G})
		
		\tcp{Sieve local excitation}
		Calculate $\hat{G}(x,y,t)$ via Eq. (\ref{Eq:G-scale},\ref{Eq:G-threshold})
		
		\tcp{Activation of membrane potential}
		Calculate $K(x,y,t)$ via Eq. (\ref{Eq:activation})
		
		\tcp{Update current dynamic threshold}
		Update $K_{thre}(t)$ via Eq. (\ref{Eq:dynamic threshold})
		
		\tcp{Spiking mechanism}
		\tcp{Output}
		Generate $\text{Spike}(t)$ via Eq. (\ref{Eq:spike})
		
		\tcp{end for}
	}
	\caption{Online algorithm}
	\label{Algorithm: inhibition model}
\end{algorithm}

\subsection{Algorithm Summary and Parameter Setting}
\begin{table}[t]
	\centering
	\caption{Parameters of the proposed inhibition model}
	\begin{tabular}{p{0.5cm}p{2cm}p{4cm}}
		\hline
		Eq. 			&Parameters  			&Description\\
		\hline
		(\ref{Eq:P})			&$N_p \in [2, 10]$ 			&persistent response\\
		(\ref{Eq:Blur}) 		&$\sigma = 1$ 				&standard deviation\\
		(\ref{Eq:FFI})  		&$C, R$ 					&columns and rows\\
		(\ref{Eq:FFI-delay})  	&$\tau = 10$		    	&time constant (ms)\\
		(\ref{Eq:FFI-delay})  	&$\Delta t \in [16, 33]$	&sampling time (ms)\\
		(\ref{Eq:GI})  			&$\beta \in [1, 10]$		&baseline sensitivity\\
		(\ref{Eq:G-scale})  	&$C_{\rho} = 0.25$			&constant coefficient\\
		(\ref{Eq:G-scale})  	&$\Delta_C = 0.01$			&small real number\\
		(\ref{Eq:G-threshold})  &$T_{g} = 2$				&local threshold gate\\
		(\ref{Eq:activation})  	&$\gamma = 0.01\ \text{or}\ 1$		&scale parameter\\
		(\ref{Eq:dynamic threshold})&$N_{dt} = N_{p}$ 		&time window\\
		\hline
	\end{tabular}
	\label{Tab: parameters}
\end{table}
Algorithm \ref{Algorithm: inhibition model} outlines the feed-forward neural computation process. 
The concomitant inhibitions operate at multiple layers and scales. 
FFI functions at the largest scale, signalling the angular size across the entire receptive field. 
In contrast to previous approaches, it undergoes non-linear mapping to interact with both SI and LI, which modulate local excitation. 
GI acts as a feedback normalization mechanism, akin to shunting inhibition, operating at the second largest scale. 
It divisively suppresses excitation in relation to its neighbouring field over time. 
SI precedes LI by curtailing local excitation initially, while LI acts with a delay, spreading inhibition to a broader area. 
These inhibitory mechanisms collectively reduce excitation engages in LGMD's activity.

Table \ref{Tab: parameters} elaborates on spatial and temporal parameters crucial for neural computation. 
Due to scaling considerations, the parameter $\gamma$ (Eq. \ref{Eq:activation}) varies in the ablation study: it is set to $1$ when GI is ablated and to $0.01$ otherwise. 
We observed that increasing the number $N_p$ of persistent photoreceptor responses improves performance in real-world scenarios, while smaller values suffice for synthetic stimuli. 
The time window of the dynamic threshold $N_{dt}$ aligns consistently with $N_p$ based on extensive data experiments, aiming to reduce false positives in complex real-world scenes. 
It is important to clarify that the proposed methods are rooted in model-based approaches aligned with anatomical and physiological findings, rather than data-driven methods. 
	Learning methods were not incorporated into the proposed inhibition model.

\subsection{Experimental Settings}
\label{Sec: model: experiment}
\begin{figure}[t]
	\centering
	\includegraphics[width=0.5\linewidth,keepaspectratio]{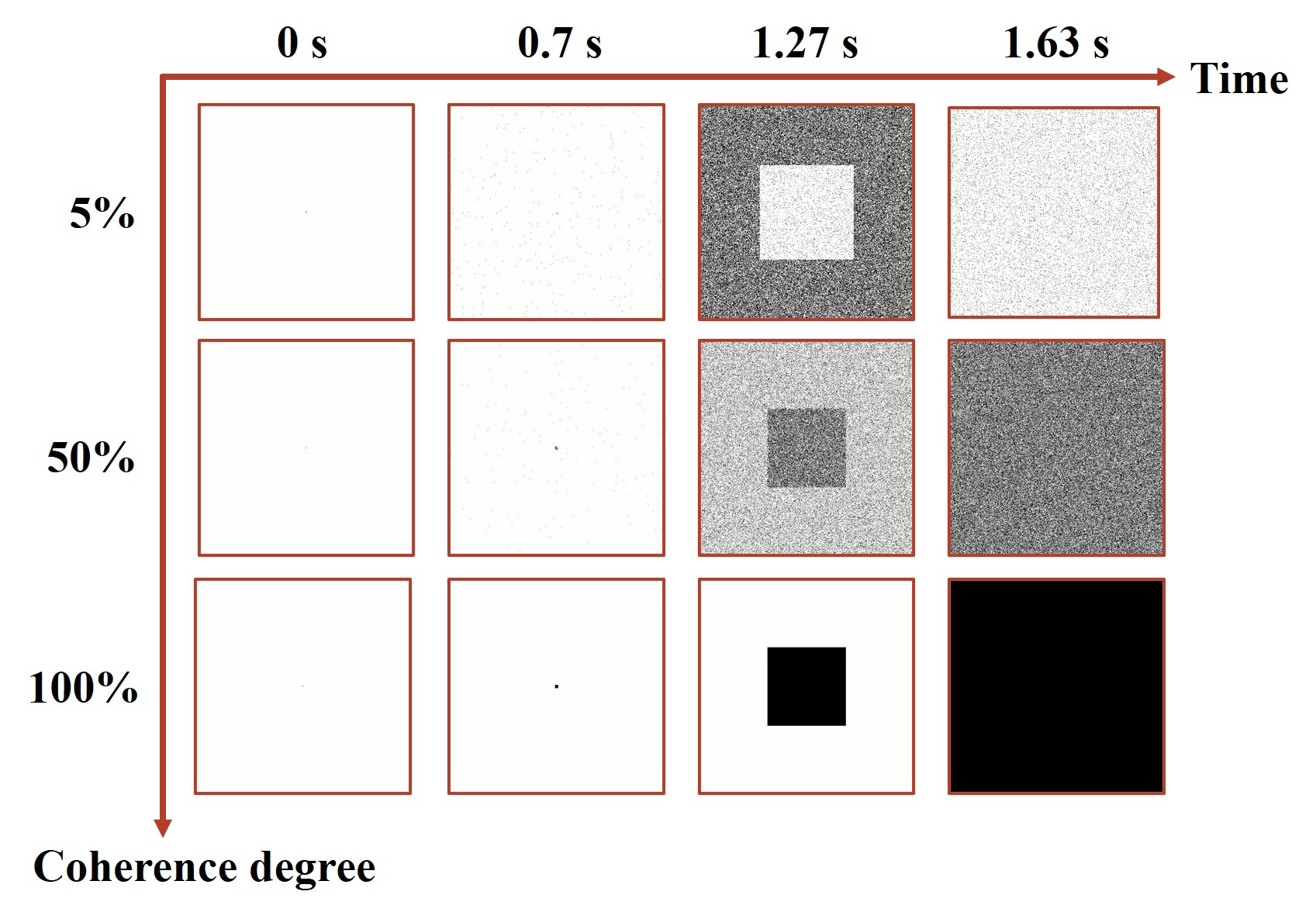}
	\caption{
		\small
		The illustrations depict coherent–incoherent dark looming stimuli within an identical time window. Here, dark looming refers to a black square expanding against a white background, representing an approaching object. The coherence degree varies from $5\%$ to $100\%$, indicating the proportion of black pixels in each frame that constitutes the looming square.
	}
	\label{Fig: coherence stimuli illustration}
\end{figure}
We conducted systematic experiments encompassing three main categories. 
Firstly, we demonstrated the foundational efficacy of our proposed inhibition model using synthetic and real-world scene stimuli. 
We statistically compared the performance of proposed model (abbreviated as "PM") with the state of the art, neural network models, specifically, the LGMD1 model in \citep{Fu-2018(LGMD1-NN)} (abbreviated as "LGMD1"), the LGMD2 model in \citep{Fu-LGMD2-TCYB} (abbreviated as "LGMD2"), the LGMD1 model in \citep{Lei-2022} (abbreviated as "LGMD1 2022 ver."), the feedback-LGMD model in \citep{LGMD-feedback-2023} (negative/positive feedback to ON channels abbreviated as "F-LGMD ONn"/"F-LGMD ONp"). 

Secondly, contrasting with previous modelling studies, we introduced coherent-incoherent motion testing across various motion patterns, comparing results with the state of the art. 
Coherence–incoherence motion is commonly employed to test neuronal responses to objects defined by either continuous or discontinuous edges. Such stimuli are not limited to studies of the LGMD; they are also widely used in primate vision research, for example, to investigate how the visual system detects figures breaking camouflage through kinetic and luminance-defined edges \citep{Layton2015}.
The coherence stimuli used in this paper were generated in a manner similar to that described in a recent study \citep{Dewell-LGMD-segmentation}. 
As shown in Fig. \ref{Fig: coherence stimuli illustration}, the 100\% coherence stimuli are the standard stimuli (also in Fig. \ref{Fig: PM basic selectivity - synthetic}), and the incoherence stimuli shares the same physical attribute (resolution and moving speed). 
These synthetic stimuli consist of dark or white squares or bars, simulating the regular motion of a solid object. 
However, in the incoherent stimuli, the pixels representing the solid object are separately and randomly distributed across the image. 
The degree of coherence defines the percentage of pixels that remain in their original positions within the moving object. 
For instance in Fig. \ref{Fig: coherence stimuli illustration}, 50\% coherence in the dark square stimuli depicts a looming process where 50\% of the "object pixels" stay in their original positions, as they would in the standard stimuli. 
The remaining 50\% of the pixels, however, are randomly distributed within the background, with their original positions in the object being replaced by background pixels. 
Importantly, the incoherent stimuli across varying degrees of coherence maintain the same overall luminance change per frame as the standard synthetic stimuli (100\% coherence). 
However, in situations where the background cannot accommodate the "object pixels", the movement process in the incoherent stimuli generation is abruptly halted, transitioning immediately to the final frames where the moving object reaches its maximum size.

Lastly, ablation studies were conducted to evaluate the effectiveness of each inhibition type in preserving specific selectivity. 
Table \ref{Tab: data collection} lists the collection of categorised experimental data tested in this study.

\begin{table}[t]
	\caption{Data collection in experimental evaluation}
	\centering
	\begin{tabular}{p{30mm}p{20mm}p{10mm}}
		\toprule[1pt]
		Category	&	Type	&	Amount	\\
		\midrule[1pt]
		\multirow{5}{=}{Synthetic stimuli}	&	\multirow{1}{=}{approach}	&	\multirow{1}{=}{48}	\\
		&	\multirow{1}{=}{recession}	&	\multirow{1}{=}{48}	\\
		&	\multirow{1}{=}{elongation}	&	\multirow{1}{=}{48}	\\
		&	\multirow{1}{=}{translation}	&	\multirow{1}{=}{48}	\\
		&	\multirow{1}{=}{grating}	&	\multirow{1}{=}{66}	\\
		\hline
		\multirow{2}{=}{Vehicle stimuli}	&	\multirow{1}{=}{crash}	&	\multirow{1}{=}{26}	\\
		&	\multirow{1}{=}{non-crash}	&	\multirow{1}{=}{10}	\\
		\hline
		\multirow{3}{=}{Indoor stimuli}	&	\multirow{1}{=}{approach}	&	\multirow{1}{=}{20}	\\
		&	\multirow{1}{=}{recession}	&	\multirow{1}{=}{20}	\\
		&	\multirow{1}{=}{translation}	&	\multirow{1}{=}{52}	\\
		\hline
		\multirow{3}{=}{Coherence stimuli}	&	\multirow{1}{=}{approach}	&	\multirow{1}{=}{22}	\\
		&	\multirow{1}{=}{recession}	&	\multirow{1}{=}{22}	\\
		&	\multirow{1}{=}{translation}	&	\multirow{1}{=}{66}	\\
		\bottomrule[1pt]
	\end{tabular}
	\label{Tab: data collection}
\end{table}

\subsection{Metric Used for Statistical Study}
To quantify the effectiveness of looming perception for all the investigated and compared models, we adopted a widely used metric indicating the accuracy of perception as the ratio of all tested data for which the model produces correct responses using the following formula
\begin{equation}
	\label{Eq:accuracy}
	\text{Accuracy} = \frac{\text{TP}+\text{TN}}{\text{TP}+\text{TN}+\text{FP}+\text{FN}} \times 100\%
\end{equation}
Precisely, TP is short for "true positives" indicating the number of tested stimulus for which the model generates correct collision alerts for genuine collision scenarios. 
This is calculated by the model releasing its initial alert to approaching objects just at or before the instant of collision. 
FP is short for "false positives" denoting the number of tested data for which the model erroneously generates collision alerts to non-colliding stimuli, such as translational or receding stimulus. 
TN, i.e., "true negatives", represents the number of tested stimulus for which the model is not responding to non-colliding stimulus. 
Lastly,
FN stands for "false negatives" accounting for the number of tested data for which the model is not responding to colliding events.


\section{Foundational Efficacy}
\label{Sec: functionality}
For the first part of experiments, we demonstrated the basic functionality of the proposed inhibition model. 
Three types of tests including synthetic stimuli testing, structured scenes testing, vehicle scenes testing were taken.

\subsection{Synthetic Stimuli Testing}
\begin{figure}[t]
	\centering
	\includegraphics[width=\linewidth,keepaspectratio]{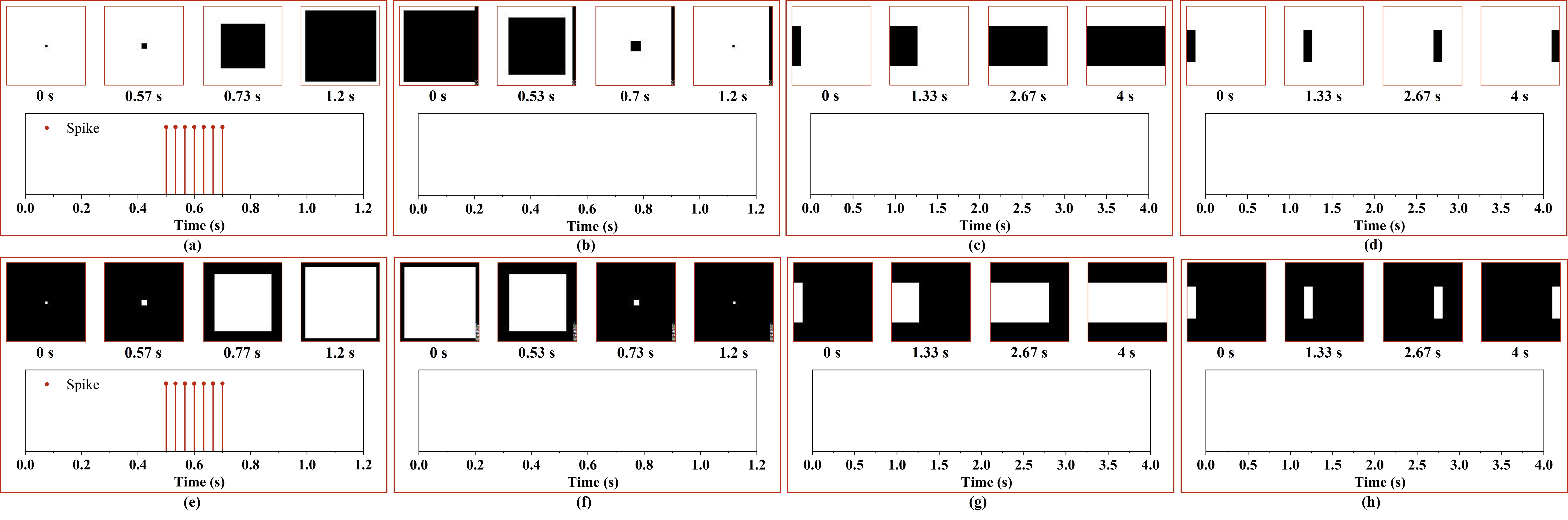}
	\caption{
		\small
		The illustrations of spiking response by the proposed model to basic motion stimuli includes 
		(a) dark square approaching, (b) dark square receding, (c) dark bar elongating, (d) dark bar translating, (e) light square approaching, (f) light square receding, (g) light bar elongating, (h) light bar translating. 
		X-axis indicates time in seconds. 
		The proposed model responds merely to dark and light approaching rather than any other kinds of movements, shows strong approaching selectivity.
	}
	\label{Fig: PM basic selectivity - synthetic}
\end{figure}
Firstly, Fig. \ref{Fig: PM basic selectivity - synthetic} shows the foundational efficacy of the proposed model under synthetic stimuli testing. 
Notably, the model selectively responds to approaching movements of either light or dark square showing successive spikes, i.e., high firing rate during approaching. 
From the time stamp in Fig. \ref{Fig: PM basic selectivity - synthetic}, we can observe that the proposed model responds to expanding square early after the beginning of motion, and is inhibited before end of movement. 
This attribute is prominent and different to previous LGMD models as the response normally gradually reaches the summit shortly before the largest angualr size subtended by moving object. 
The rationale behind is the efficacy window of the concomitant inhibitions. 
SI curtails local excitation quickly in situ at the beginning of approaching, and at this moment, LI is weak. 
The slight delay makes SI effects inner to expanding edges so the local excitation can build up to activate the LGMD. 
As the angular size increases, i.e., consistently expanding of object, SI becomes trivial while LI suppresses the spreading-out excitation more strongly. 
Accordingly, the model can predict collision from expanding of small-size target.

\begin{figure}[t]
	\centering
	\includegraphics[width=0.65\linewidth,keepaspectratio]{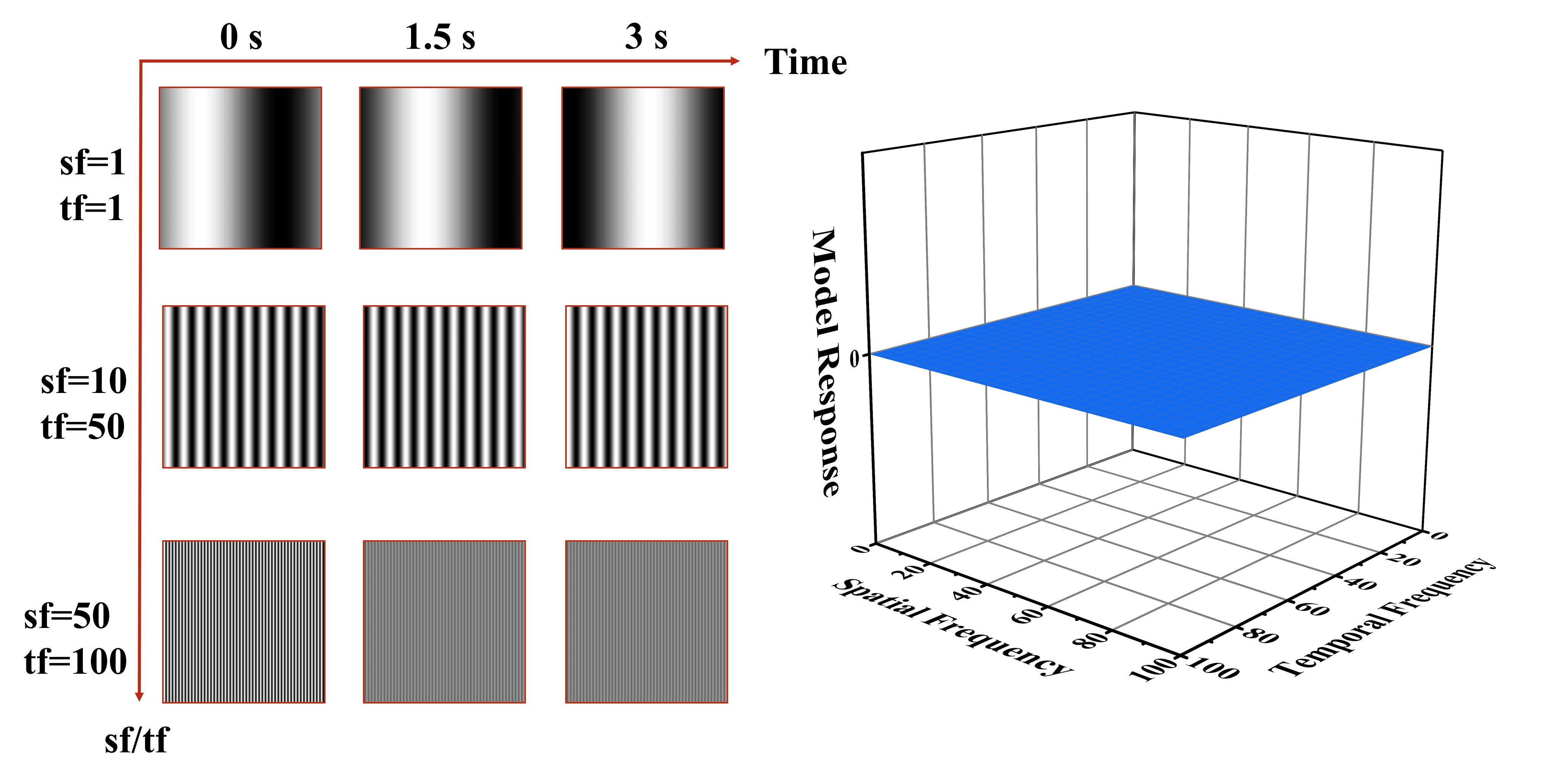}
	\caption{
		\small
		The illustration of the proposed model response by grating stimuli at a variety of spatial and temporal frequencies: 
		the model is inactive against all grating movements.
	}
	\label{Fig: PM basic selectivity grating}
\end{figure}

During dark or light square recession, LI and SI effects subsequently to suppress the neuronal activities. 
Challenged by elongation and translation, SI works effectively to cancel local excitations, consistent with the biological hypothesis \citep{LGMDs-2016}. 
Likewise, we carried out sinusoidal grating testing under a wide range of spatial and temporal frequencies. 
As shown in Fig. \ref{Fig: PM basic selectivity grating}, the proposed model remains inactive to all tested grating visual stimuli, a critical property of looming perception neuronal model in harmony with previous studies upon LGMD computational modelling (see review in \citep{Fu-ALife-review,Fu-ON/OFF-2023}).

\subsection{Real World Stimuli Testing}
The real world tests including structured indoor visual stimuli induced by ball movements, and complicated vehicle crash scenes constructed by collecting dashboard camera recording, adapted from a dataset \citep{Fu-2020-Access}. 
The former class includes recording frames of proximity, recession, and translation of dark or white balls, which correspond to the synthetic motion patterns introduced earlier (i.e., approach, recession, and translation). The latter class involves real-world scenarios of car crash and non-crash events.

\begin{figure}[t]
	\centering
	\includegraphics[width=\linewidth,keepaspectratio]{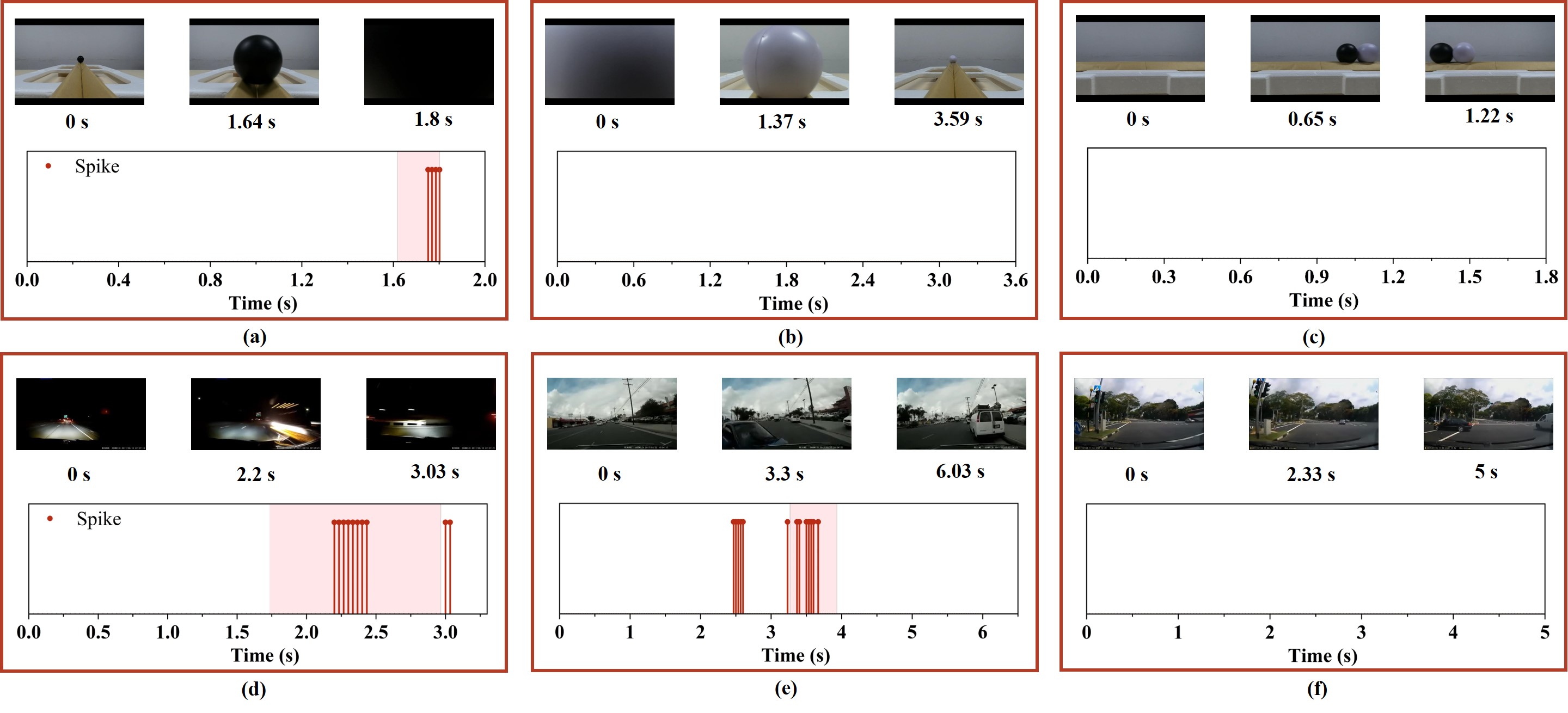}
	\caption{
		\small
		The illustration of spiking response of the proposed model under real world stimuli testing -- the shaded area indicates ground-truth time window for collision alert: 
		(a) dark ball approaching, 
		(b) white ball receding, 
		(c) dark and white balls translating, 
		(d) car crash at night, 
		(e) car crash at day time, 
		(f) car non-crash event.
	}
	\label{Fig: PM selectivity realworld}
\end{figure}

Firstly, Fig. \ref{Fig: PM selectivity realworld} represents the model performance against a variety of real physical visual stimuli. 
The proposed inhibition model spikes for imminent collisions in both indoor scene and vehicle navigation, whilst remains silent to receding, translating stimuli induced by ball movements. 
The model works effectively to predict collision danger with high firing rates within the indicated critical time window before real collision. 
Particularly, the model could respond early to real crash as demonstrated by Fig. \ref{Fig: PM selectivity realworld}(e). 
Consequently, the proposed method shows robust selectivity to only approaching movements.

\begin{figure}[t]
	\centering
	\includegraphics[width=\linewidth,keepaspectratio]{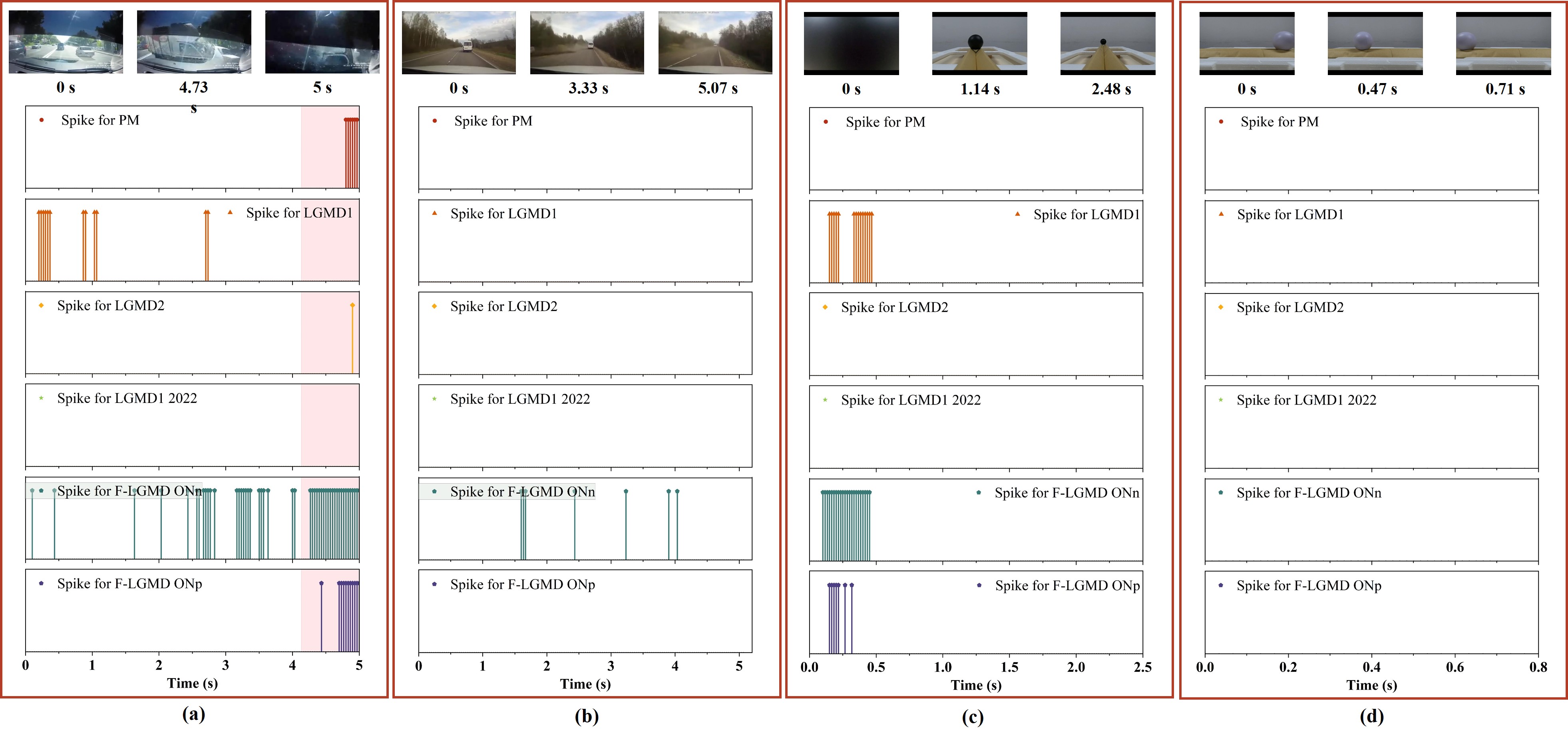}
	\caption{
		\small
		The illustration of spiking response of the proposed model and comparative models under real world stimuli testing: 
		(a) car crash, 
		(b) normal navigation,
		(c) dark ball receding,
		(d) white ball translating. 
	}
	\label{Fig: Compare2sota - realworld}
\end{figure}

\begin{figure}[t]
	\centering
	\includegraphics[width=0.5\linewidth,keepaspectratio]{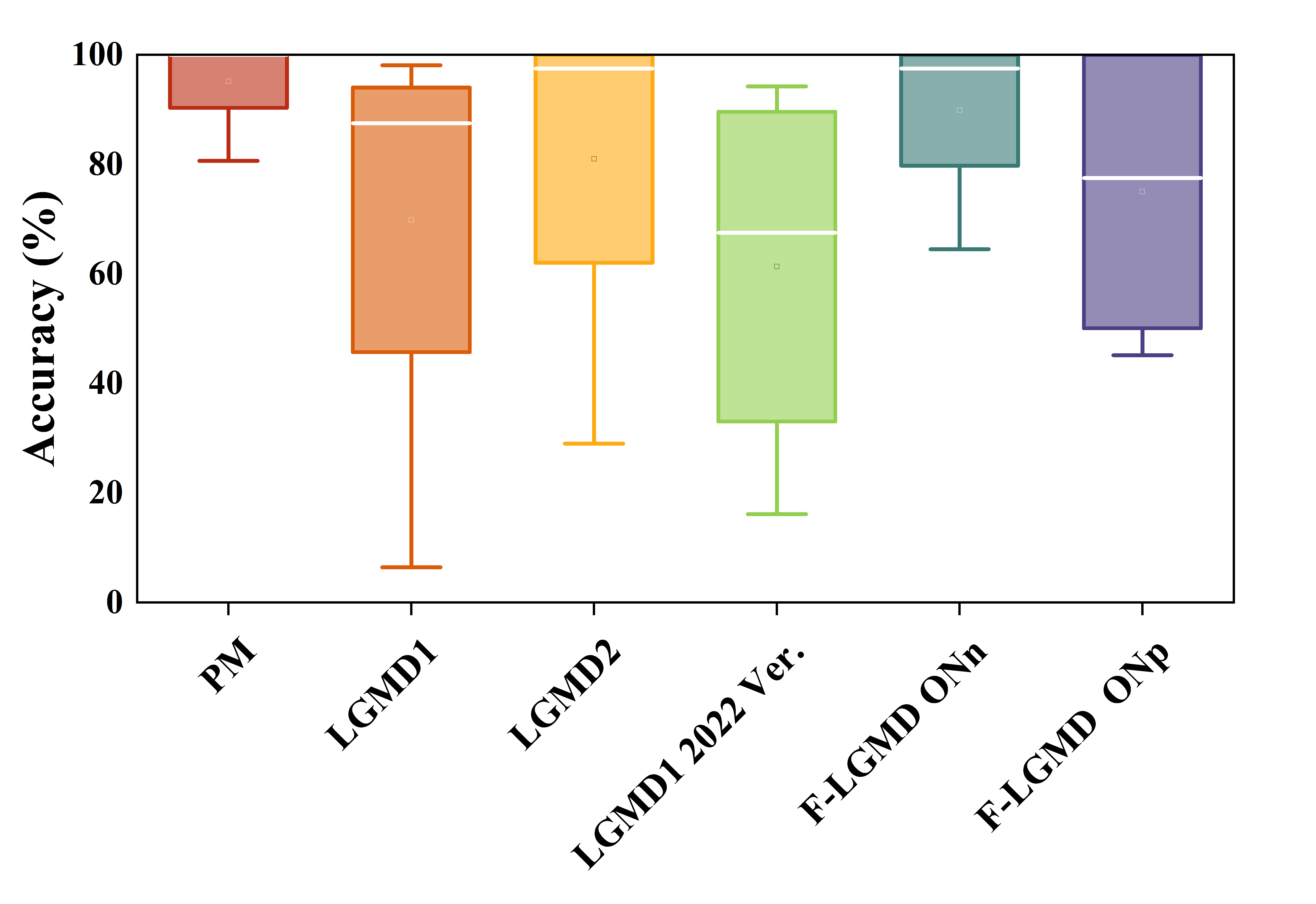}
	\caption{
		\small
		The statistical results of comparative experiments against real world visual challenges: the median and variance indicate the accuracies across five categories of stimuli, i.e., ball-approach, ball-recession, ball-translation, vehicle-crash, vehicle-non-crash. 
		The accuracy is computed via Eq. \ref{Eq:accuracy}. 
		The tested data is listed in Table \ref{Tab: data collection}.
	}
	\label{Fig: Compare2sota accuracy boxplot}
\end{figure}

To further corroborate the improvements by the proposed modelling of concomitant inhibitions, we also conducted comparative experiments with the state of the art, using the dataset shown in Table \ref{Tab: data collection}. 
Fig. \ref{Fig: Compare2sota - realworld} showcases the comparative models with the proposed model with respect to spiking response to vehicle and indoor visual movements including crash, normal road navigation (non-crash), recession, and translation occurrences. 
The results demonstrate the proposed model responds to collision and other kinds of movements, accurately. 
In this regard, the LGMD2 model performs also well to the challenges. 
However, the LGMD1 model shows false positives in vehicle crash and ball recession scenarios. 
The LGMD1 2022 ver. model cannot predict car crash danger, the model performance of which is rigorously restricted by its signal competition between ON/OFF channels. 
The F-LGMD ONn, and F-LGMD ONp models nevertheless are not competitive to the proposed method.

Moreover, Fig. \ref{Fig: Compare2sota accuracy boxplot} compares statistically the models challenged by the synthetic and real-world dataset in Table \ref{Tab: data collection}. 
The median accuracy of the proposed model approaches closely to $100\%$ with the smallest variance across the whole real world dataset. 
The statistical results verified the superiority of proposed model compared with the state of the art, preserving the "approaching selectivity" across various scenarios involving complex and dynamic environments.


\section{Coherent-Incoherent Motion Testing}
\label{Sec: coherence}

\begin{figure}[t]
	\centering
	\includegraphics[width=\linewidth,keepaspectratio]{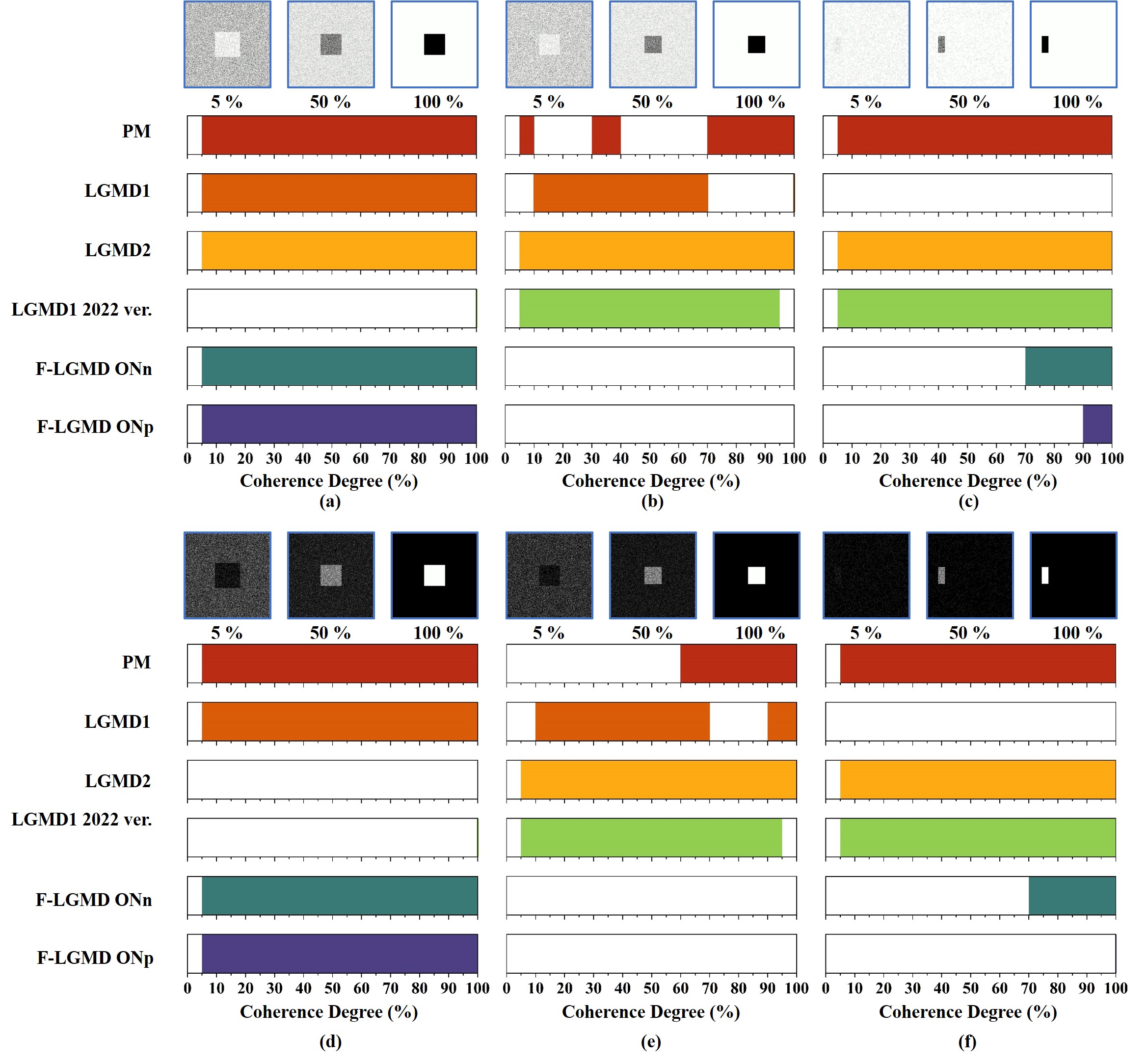}
	\caption{
		\small
		The illustration of coherence-incoherence stimuli testing of all comparative LGMD models against dark and light approach, recession, and translation stimuli, at degree of coherence varying between $5\%$ and $100\%$ for each kind of stimuli: the shaded area indicates respective perception accuracy (TP$+$TN). 
		(a) dark proximity, (b) dark recession, (c) dark translation, (d) light proximity, (e) light recession, (f) light translation.
	}
	\label{Fig: Compare2sota coherence stimuli}
\end{figure}

For the second part of experiments, we compared and investigated the performance of models against coherence-incoherence motion stimuli. 
Previous computational studies lacked systematic examination on incoherence stimuli whereas recent physiological studies pointed out looming sensitive neurons, like the LGMD, demonstrate spatial selectivity \citep{Dewell-LGMD-segmentation,Contrast-polarity-mapping-2022}. 
An object on a collision course will generate a specific sequence of images representing expanding edges across receptive field. 
These images will gradually activate LGMD's dendrites in a specific pattern. 
Such a pattern can be spatially formulated either continuously, i.e., coherence, or scrambled, i.e., incoherence.

As presented in Section \ref{Sec: model: experiment}, Fig. \ref{Fig: coherence stimuli illustration} exemplifies dark looming squares at the lowest, the intermediate, and the highest coherence degree where the highest degree corresponds to the stimulus used in the synthetic stimuli testing in Section \ref{Sec: functionality}. 
These movies, at same physical attribute yet different coherence degrees, thus activate the investigated models in different motifs. 
The biological LGMD neurons responded more to the non-scrambled, coherent visual stimuli. 
As of yet, we expect the computational models can recognise approaching squares even at very low coherence degree. 
This could significantly benefit the use in complex and dynamic visual tasks like self-driving cars, or in technology to help visually impaired people navigate independently.

Fig. \ref{Fig: Compare2sota coherence stimuli} demonstrates perception accuracy (TP$+$TN) of all comparative models. 
Firstly, regarding approach case, all comparative models can fully detect approaching squares at the intact range of tested coherence degree ($100\%$ TP), whereas the LGMD1 2022 ver. model with competitive ON/OFF channels can detect merely $100\%$-coherence looming. 
In particular, the LGMD2 model possesses the specific selectivity to only darker objects approaching, briefly responds to lighter receding objects, in consistent with its physiological characteristics \citep{Simmons-1997(LGMD2-neuron-locusts)}. 
Accordingly, the LGMD2 model is not a proper model system dealing with lighter approaching as indicated in Fig. \ref{Fig: Compare2sota coherence stimuli}(d).

Secondly, the movements induced by an object moving away from the receptive field are always challenging artificial collision-detecting vision systems whereas the biological LGMD neurons can discriminate such motion patterns well representing brief or none firing rate. 
We always expect the looming perception models not responding to receding stimulus even at the start of recession. 
Our proposed model nevertheless is not competitive to all tested coherence degrees. 
The rational behind could be the lateral inhibition is not strong enough to shunt down the LGMD immediately. 
We will look deeper into the effect of LI regarding recession case in ablation studies. 
On the other hand, the LGMD1 2022 ver. model, and the LGMD2 model outperformed other candidates with regard to recession testing where the LGMD2 shows $100\%$ TN. 
In contrast, the F-LGMD ONn, and F-LGMD ONp models demonstrate the highest FN.

Lastly, when challenged by coherence-incoherence translation stimuli, the proposed model, the LGMD2 model, the LGMD1 2022 ver. model showed competitive robustness at $100\%$ TN. 
Being different from these two models, we proposed the modelling of self-inhibition that works effectively to suppress translating motion, in harmony with physiological research \citep{LGMDs-2016}. 
Moreover, the LGMD1 2022 ver. model formulated neural competition between ON/OFF channels to achieve the antagonism to translation which, however, has higher computational complexity. 
Conversely, the proposed model can simply implement such selectivity regardless of encoding ON/OFF-contrast.


\section{Ablation Study}
\label{Sec: ablation}
In the previous parts of experiments, we investigated and compared the proposed inhibition model with the state of the art. 
Those results not only demonstrated the superiority of this method with regard to robust selectivity towards only approaching motion, but also articulated the efficacy of modelling multi-scale, multi-level inhibitions in close concert with biological theories. 
Here we provide comprehensive studies through ablating inhibitory signal processing, separately and in succession, to look into their respective efficacy.

\subsection{Ablating Global Inhibition}
GI acts as instant, feedback normalisation of light intensities into neighbouring fields upstream of SI and LI. 
Here we ablated GI via skipping such processes of Eq. (\ref{Eq:GI},\ref{Eq:GI-conv}), then conducted contrast-specific tests.

\begin{figure}[t]
	\centering
	\includegraphics[width=0.5\linewidth,keepaspectratio]{Fig12.jpg}
	\caption{
		\small
		The illustration of contrast-specific tests: the radar graph indicates perception accuracy (TP$+$TN) against synthetic dark and light square approaching (DA, LA), receding (DR, LR), elongating (DE, LE), and translating (DT, LT) at contrast between moving object and background varying from 20 to 255 in greyscale. 
		(a) Accuracy of the proposed model with GI: the contrast boundary details are amplified in the inset. 
		(b) Accuracy after ablating GI: the accuracy to dark/light receding movements is abolished and the contrast boundary to approach is smaller.
	}
	\label{Fig: Ablation without GI contrast}
\end{figure}

Fig. \ref{Fig: Ablation without GI contrast} elaborates on perception accuracy of the proposed inhibition model under a variety of synthetic stimuli. 
Through the tests, we found the performance boundaries of the proposed model to contrast-varied stimuli. 
It turns out that GI significantly impacts on the proposed model, regarding both selectivity and contrast responsiveness. 
The model response to receding stimuli cannot be eliminated, as the GI naturally reduces the excitation at expanding edges via normalising it into its neighbouring field; this effect is more prominent with high-contrast movements. 
The contrast boundary becomes smaller after ablating GI, with the lowest from 25 to 50 (the intensity differences between foreground moving object and static background). 
The results demonstrated GI is capable of increasing the fidelity of low-contrast looming perception, in consistent with recent studies \citep{Drews-dynamic-signal-compression,Fu-IJCNN-2021,Fu-Array}.

\subsection{Ablating Feed-forward Inhibition}
\begin{figure*}[t]
	\centering
	\includegraphics[width=\linewidth,keepaspectratio]{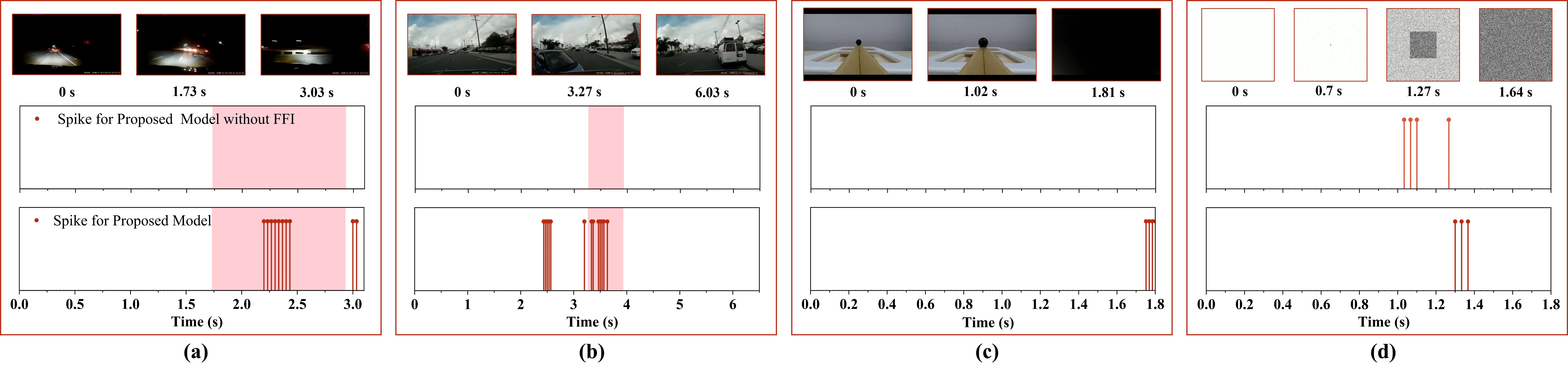}
	\caption{
		\small
		The spiking response of the proposed model to looming stimuli under previously tested scenes after (second row), and before (third row) ablating FFI: the shaded area in vehicle testing results indicates ground-truth time window for collision alert. 
		(a) vehicle crash at night, (b) vehicle crash in the daytime, (c) ball approach, (d) incoherent dark approach. 
		Ablating FFI abolishes the approaching selectivity.
	}
	\label{Fig: Compared-FFI}
\end{figure*}
In typical LGMD neural network models, FFI obeys to an all-or-none law that can directly shunt down the neuronal activities once brightness changes rapidly over a large area of receptive field \citep{Fu-ALife-review}. 
Later studies endeavoured to make FFI more flexible, transformed the FFI output to time-varying coefficient with LI which improves the robustness of model in complex dynamic scenes, e.g., vehicle scenarios, moreover preserves the selectivity \citep{Fu-2020-Access,Fu-LGMD2-TCYB}. 
In this work, the FFI is likewise computed from average luminance change, while interacting with LI and SI to yield a complementary modulation that suppresses feed-forward excitation presynaptic to the LGMD.
As the FFI can indicate angular size, the motivation herein is matching the physiological findings that effects of SI and LI in visual neural systems are changed by the temporally varied angular size.

Fig. \ref{Fig: Compared-FFI} articulates the spiking response of the proposed model without and with FFI. 
It can be clearly seen from the results that if ablating FFI, the proposed model no longer demonstrates the selectivity to approaching motion, the basic attribute, especially in real-physical visual scenes. 
More specifically, in this case, the coefficients with SI and LI were equalised rather than computed via Eq. (\ref{Eq:AS}), i.e., $\omega(t) \equiv 0.5$. 
Too strong self-inhibition will immediately suppress local excitation in situ, the LGMD thus is hard to receive any excitatory signals from its dendrite. 
Accordingly, in our proposed looming perception model, FFI decides the basic selectivity, which should be computed with respect to time.

\subsection{Ablating Self-Inhibition}
SI is a newly modelled type of inhibition in this research. 
We would anticipate the effect of SI reconciles with the physiological studies inspired us. 
SI acts with a short delay to cut down excitation in situ, before it spreading out in space. 
LI could be formed in cascade by such spread-out excitation. 
Accordingly, SI in concert with LI interact with excitation at multiple time and space scales, formulated in multi-layers of neural network as depicted in Fig. \ref{Fig: four inhibitions}. 
By ablating SI in the model, we skipped the neural computation in Eq. (\ref{Eq:SI-convolve},\ref{Eq:SI-delay}).

\begin{figure}[t]
	\centering
	\includegraphics[width=\linewidth,keepaspectratio]{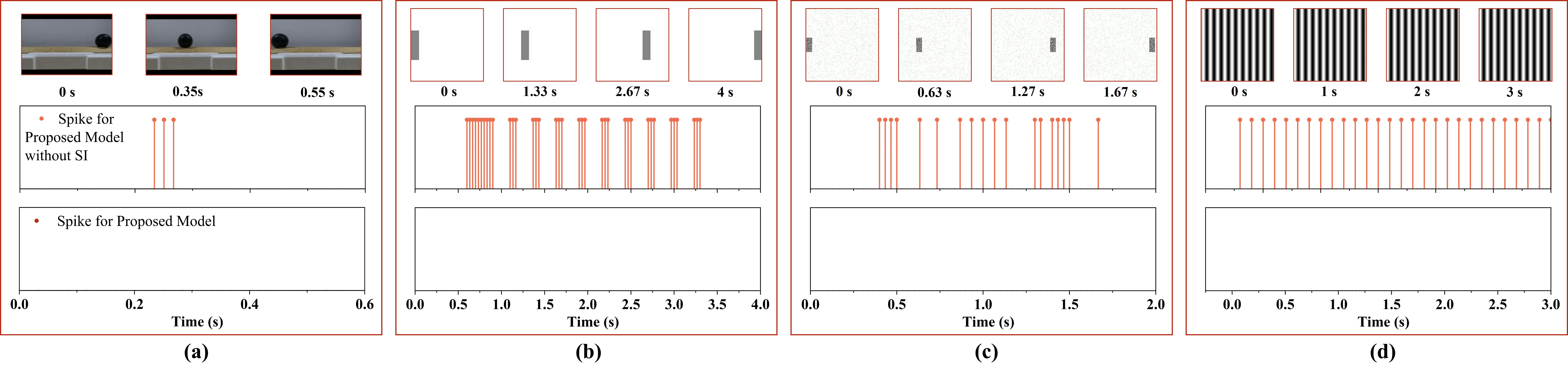}
	\caption{
		\small
		The spiking response of the proposed model to translating and grating stimuli under previously tested scenes after (second row), and before (third row) ablating SI: 
		(a) ball translation, (b) synthetic dark translation, (c) incoherent dark translation, (d) sinusoidal grating. 
		Ablating SI abolishes the antagonism to translating stimuli.
	}
	\label{Fig: Compared-SI}
\end{figure}

We can clearly see from results in Fig. \ref{Fig: Compared-SI} that ablating SI abolishes the antagonism to translating motion including grating. 
Grating movements would induce high angular size that are suppressed by the coordination of SI and LI. 
As demonstrated in Section \ref{Sec: functionality}, SI cancel out response by translation very effectively as the proposed model showed $100\%$ TN challenged in every kind of translational stimuli tests. 
Therefore, our proposed computational model shows a convincing way of numerically simulating SI to fight against translational-like motion.

\begin{figure}[t]
	\centering
	\includegraphics[width=\linewidth,keepaspectratio]{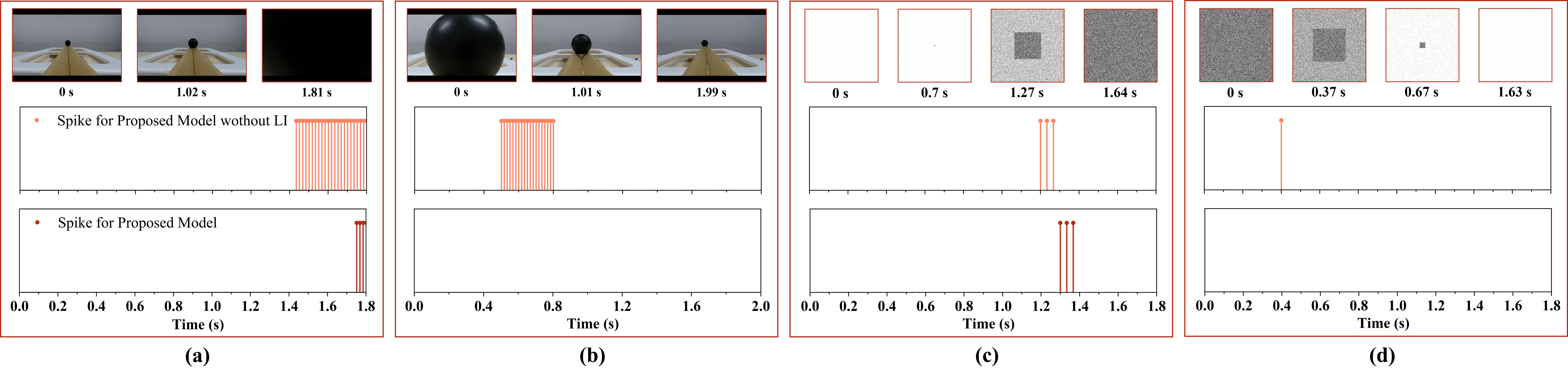}
	\caption{
		\small
		The spiking response of the proposed model to approach and recession stimuli under previously tested scenes after (second row), and before (third row) ablating LI: 
		(a) ball approach, (b) ball recession, (c) incoherent dark approach, (d) incoherent dark recession. 
		Ablating LI cannot eliminate the model response to recession.
	}
	\label{Fig: Compared-LI}
\end{figure}

\subsection{Ablating Lateral Inhibition}
Next, we kept GI, FFI, SI, and ablated LI via skipping the neural computation in Eq. (\ref{Eq:SI-passing},\ref{Eq:LI-convolve},\ref{Eq:LI-delay}). 
In this regard, SI is placed as the only local type of inhibition reducing excitation. 
The computation of coefficient with SI was also associated with FFI. 
Concretely, the increase of angular size would lead to dramatic decrease of the coefficient with SI, and thus the neuronal activities are hard to be suppressed at the moment of moving target subtending large angular size.

Fig. \ref{Fig: Compared-LI} consolidates our thinking that ablating LI could make the model no longer capable of discriminating recession from approach though the spiking frequency during recession could be lower than that during approach. 
The results herein are in line with previous modelling studies, as LI hitherto has undergone decades of research into the forming of looming selectivity since 1990s \citep{LGMD1-1996(Rind-intracellular-neurons),LGMD1-1996(Rind-neural-network),LGMD1-1999(Rind-seeing-collision)}.

\begin{figure}[t]
	\centering
	\includegraphics[width=\linewidth,keepaspectratio]{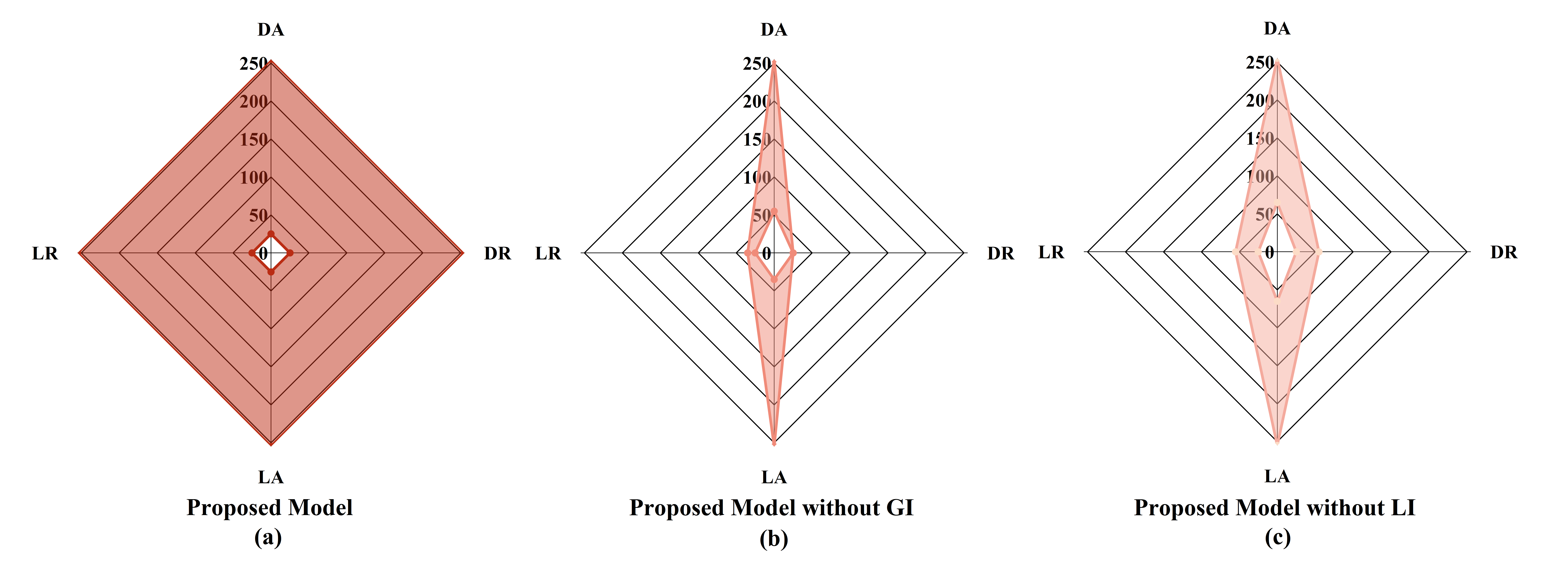}
	\caption{
		\small
		The illustration of contrast-specific tests: the radar graph indicates perception accuracy (TP$+$TN) against synthetic dark and light square approaching (DA, LA), and moving away (DR, LR). 
		(a) Accuracy of the proposed model with intact inhibitory signal processing: results are in accordance with Fig. \ref{Fig: Ablation without GI contrast}a. 
		(b) Accuracy after ablating GI: the results align with Fig. \ref{Fig: Ablation without GI contrast}b. 
		(c) After ablating only LI, the accuracy to recession stimuli is greatly attenuated.
	}
	\label{Fig: Ablation without GI_LI contrast compare}
\end{figure}

As the ablation study upon GI has already demonstrated very high FN of the proposed model (Fig. \ref{Fig: Ablation without GI contrast}), we established a new finding similarly to the ablation of LI. 
We also conducted contrast-specific tests by removing LI. 
The results in Fig. \ref{Fig: Ablation without GI_LI contrast compare} can be abstracted as two main points: 
(1) in the proposed model, removing either GI or LI would greatly attenuate the TN of the proposed model against recession stimuli; 
(2) GI still increases the contrast boundary responded properly.

\begin{figure}[t]
	\centering
	\includegraphics[width=0.6\linewidth,keepaspectratio]{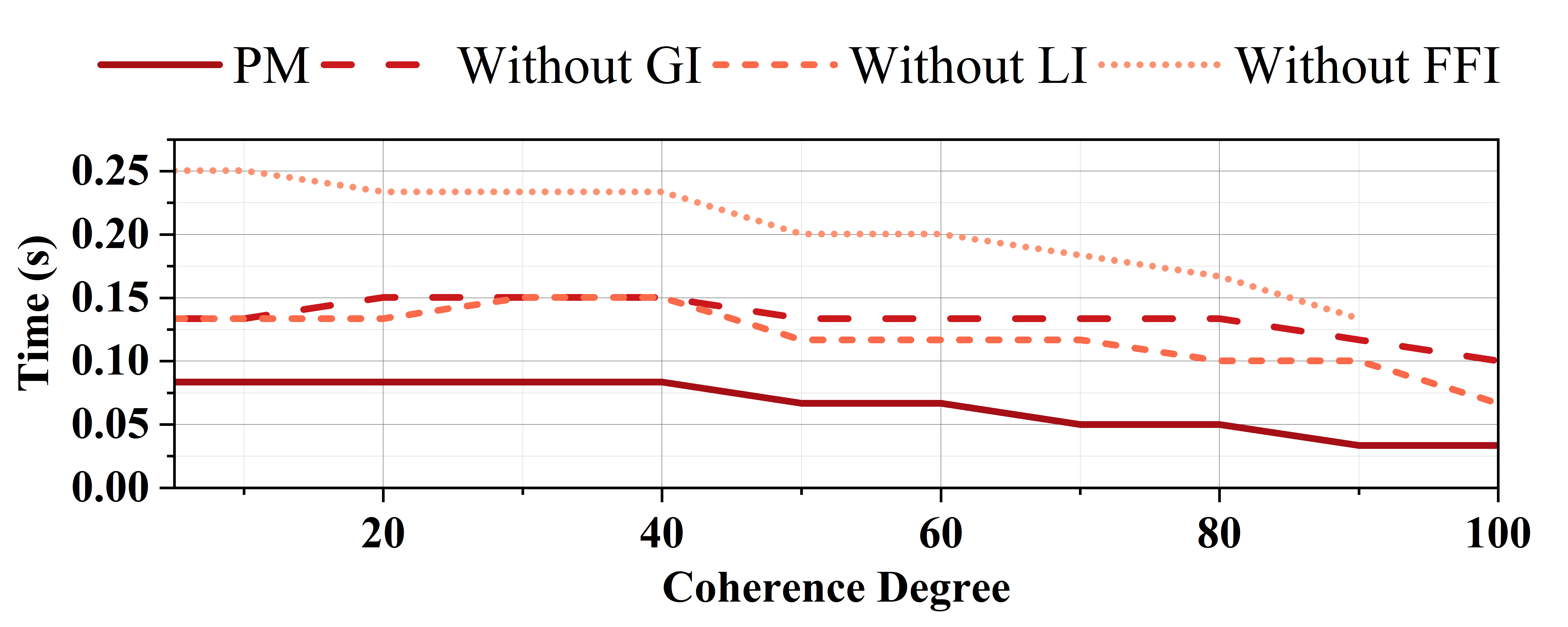}
	\caption{
		\small
		The illustration of TTC results challenged by coherent-incoherent approach at all tested coherence degrees: 
		the model with all types of inhibition demonstrates the lowest TTC.
	}
	\label{Fig: 2dcoherenceTOC}
\end{figure}

\subsection{Time to Contact}
Finally, we investigated an imperative property of collision perception visual systems, i.e., time to contact (TTC), that can indicate the time difference between the summit of neuronal firing rate and the genuine collision moment. 
In this study, the TTC is likewise obtained by counting the total discrete frames in seconds between the initial spike elicited by LGMD and the maximum angular size of looming square. 
We applied the coherence-incoherence approaching stimuli to investigate the TTC through ablating certain type of inhibition.

Fig. \ref{Fig: 2dcoherenceTOC} compares the intact model, and the model via ablating GI, LI, FFI, respectively. 
As the main impact of SI is intrinsically on suppressing translation-typed motion and it has little influence on approaching movements, the model without SI was precluded in this kind of comparative experiments. 
Intuitively, the intact model predicts collision most precisely, showing the smallest TTC across all tested coherence degrees. 
The models ablated GI and LI show very similar TTC, at intermediate levels. 
The model ablated FFI exhibits the largest TTC, even cannot recognise looming squares at coherence degrees greater than $90\%$.


\section{Further Discussions}
\label{Sec: discussion}

The natural world always surprises us with powerful sources. 
Biological systems are robust and good inspirations for building artificial vision systems. 
The bio-inspired models have certain advantages over data-driven methods due to their computational simplicity and efficiency, etc. 
Biologists have identified inhibitory neurons that encode the angular size of simulated objects on a collision course, underscoring approaching object perception as a fundamental capacity for survival. 
This shows that inhibitory neurons provide time-varying information on spatially extended stimuli and contribute to neuronal computations involved in collision avoidance behaviours. 
However, the pre-synaptic neural networks converging different inhibitory signals and their role in shaping the specific looming selectivity remain poorly understood. 
This modelling study stemmed from related biological system models, and provided new insights into how we could formulate the different types of inhibition to be combined in pre-synaptic visual processing of a looming sensitive neuron, which would be a solid step toward bio-inspired intelligence for collision detection.

\subsection{Reflections on the Proposed Inhibition Model}
Similar to previous biological research on global inhibition's functionality \citep{Zhu-LGMD-lateral-excitation}, the global inhibition in our model scales the neuronal response within a certain range regardless of the object-background contrast, thereby enhancing the looming selectivity of the model under low-contrast conditions. 
Prior to this study, Olson's work also numerically modelled global inhibition by normalising local signals with whole-field signals, ensuring that the neuronal output remains within a consistent range \citep{Feedforward-global-lateral-inhibition-model}. 
Unfortunately, contrast-related research was not addressed in their work.

Lateral inhibition has been validated and numerically modelled since Rind's seminal work in 1996 \citep{LGMD1-1996(Rind-neural-network)}. 
In contrast, self-inhibition was recently proposed by Rind in 2016 \citep{LGMDs-2016}, but it has not yet been further validated. 
We are the first to numerically model self-inhibition, inspired by Rind's work. 
self-inhibition occurs in TmAs and interacts reciprocally within a shorter time window than lateral inhibition, leading us to ideally assume that self-inhibition precedes lateral inhibition and operates on a smaller scale. 
It is worth noting that previous biological research suggested that lateral inhibition might occur at a later stage of neural signal processing in the LGMD neuron, which implies that self-inhibition and lateral inhibition could potentially overlap during the processing stages. 
Therefore, in the proposed model, we combine self-inhibition and lateral inhibition, with this combination regulated by feed-forward inhibition mechanism in a complementary fashion.

In the proposed model, self-inhibition effectively suppresses the model's response to translating-like stimuli, including whole-field gratings. 
Meanwhile, lateral inhibition enables the model to distinguish between approaching and receding stimuli. 
However, these findings differ from previous biological research, which suggested that it is lateral inhibition that suppresses the LGMD's response to gratings. 
This discrepancy leads us to hypothesize that self-inhibition may transition into lateral inhibition under certain conditions, or that their functional roles dynamically shift depending on stimulus context.  
Additionally, through numerical modelling of these inhibitions, we observed that the absence of lateral inhibition does not affect the model's ability to distinguish between approaching and receding stimuli under low-contrast conditions. 
This may be due to the low excitation levels of the afferent input signal in such stimuli. 
Since low excitation results in even lower inhibition within the proposed model, the absence of this minimal inhibition does not significantly impact the model's response.

These findings resonate with psychophysical studies on induced motion, which demonstrate that the perception of translational motion can be confounded by depth-related cues \citep{Leveille2014,Leveille2010}. Such observations provide external support for the division of roles identified in our model: self-inhibition suppresses translation-like responses that might otherwise be confounded with depth cues, whereas lateral inhibition operates on depth-related signals to differentiate approach from recession.

Beyond these local mechanisms, feed-forward inhibition in the proposed model operates across the entire receptive field, reflecting the physical attribute of the angular size of a moving target. 
This approach is consistent with biological research on feed-forward inhibition in the LGMD neuron and its afferent network. 
Unlike previous LGMD models that treated feed-forward inhibition as an all-or-none trigger, the feed-forward inhibition in this study serves as an angular size indicator, facilitating the regulation of self-inhibition and lateral inhibition. 
This configuration significantly enhances the looming selectivity of the proposed model across complex testing environments. 
However, it is important to note that most neuroscience research suggests that feed-forward inhibition might be threshold-triggered, a consideration not accounted for in this study.

\subsection{Prevalence of Inhibition in Motion Perception}
Inhibition has also been shown to preserve different motion selectivities in dragonfly and fruit fly \textit{Drosophila}. 
In the visual systems of the dragonfly and \textit{Drosophila}, there are object detection neurons with specific size selectivity, such as small target movement detecting (STMD) neurons \citep{Dragonfly-2012(visual-prey-capture), hennig2008distributed, warzecha1993neural}. 
These neurons have distinct optical width preferences, responding to angular sizes of only a few degrees within their receptive fields \citep{Fu-ALife-review}. 
Inhibition has been demonstrated to spatially confine neuronal selectivity to the size of moving objects \citep{geurten2007neural, bolzon2009local, barnett2007retinotopic}.

Fly visual systems process optical flow through T4 and T5 cells and their downstream lobula plate tangential cells (LPTCs) \citep{Borst-Review-2019-Fly,Borst-review-2023,Borst2011(review-motion),Borst2015(common-circuit-motion)}. 
LPTCs are naturally stratified according to distinct directions and integrate local responses from T4 (ON) and T5 (OFF) cells \citep{Maisak_2013(T4-T5-fly)}. 
To understand how strong direction-selective responses are generated and maintained in complex natural scenes, biologists discovered inhibitory connections between adjacent sub-layers of LPTCs, specifically lobula plate-intrinsic (LPi) neurons \citep{Mauss-Opposing-Motions-Fly,Badwan-Dynamic-nonlinearities-EMD}. 
The inhibitory LPi neurons contribute to the direction-opponent responses of wide-field LPTCs, enhancing flow-field selectivity during lift and forward flight. 
Additionally, LPi neurons shape the ultra-selectivity of a specific ensemble of visual projection neurons, such as LPLC2, to expanding objects at the centre of view, with opponency to radial motion \citep{LPLC-Nature-2017}.


\section{Conclusion}
\label{Sec: conclusion}
This paper proposed a new perspective of modelling and associating four types of inhibitions consistently in a neural network framework, which were investigated most separately in previous studies. 
These inhibitions act effectively to interact with excitatory signals pre-synaptic to the looming selective LGMD, at multiple space and time scales. 
FFI works across the whole receptive field, indicating the physical attribute of angular size subtended by moving target. 
A notable attempt in this study is such an indicator is transformed to form a 
complementary integration of self-inhibition and lateral inhibition interacting with local excitation. 
This aimed at portraying the revealed physiological phenomenon that lateral inhibition could be gradually more effective along with the increase of angular size. 
The opposite holds true for self-inhibition, which was for the first time numerically formulated and incorporated with lateral inhibition. 
In harmony with physiological findings, self-inhibition works most effectively to curtail excitation induced by translational-like motion. 
In addition, global inhibition functions upstream of self-inhibition and lateral inhibition to normalise local excitations which increases the contrast boundary of looming perception. 
Working together, the proposed model is capable of responding only to approaching motion even in complex and dynamic real world scenarios. 
The systematic experiments not only demonstrated the emergence of robust approaching selectivity through neural computation of concomitant inhibitions naturally existing in biological neural systems, but also elucidated the computational role of each typed inhibition.


\section*{Acknowledgments}
This research was supported by the National Natural Science Foundation of China under Grant No. 62376063.

\bibliographystyle{elsarticle-num}

\bibliography{qinbingbib.bib}

\begin{thebibliography}{10}
\expandafter\ifx\csname url\endcsname\relax
  \def\url#1{\texttt{#1}}\fi
\expandafter\ifx\csname urlprefix\endcsname\relax\def\urlprefix{URL }\fi
\expandafter\ifx\csname href\endcsname\relax
  \def\href#1#2{#2} \def\path#1{#1}\fi

\bibitem{LGMD-1974}
M.~O'Shea, J.~L. Williams, The anatomy and output connection of a locust visual
  interneurone; the lobular giant movement detector ({LGMD}) neurone, Journal
  of Comparative Physiology 91~(3) (1974) 257--266.

\bibitem{Rind1998(local-circuit-locust)}
F.~C. Rind, P.~J. Simmons, Local circuit for the computation of object approach
  by an identified visual neuron in the locust, Journal of Comparative
  Neurology 395~(3) (1998) 405--415.

\bibitem{LGMD1-1999(Rind-seeing-collision)}
F.~C. Rind, P.~J. Simmons, Seeing what is coming: Building collision-sensitive
  neurones, Trends in Neurosciences 22~(5) (1999) 215--220.

\bibitem{LGMD1-1996(Rind-intracellular-neurons)}
F.~C. Rind, Intracellular characterization of neurons in the locust brain
  signaling impending collision, Journal of Neurophysiology 75~(3) (1996)
  986--995.

\bibitem{Borst-Review-2019-Fly}
A.~Borst, J.~Haag, A.~S. Mauss, How fly neurons compute the direction of visual
  motion, Journal of Comparative Physiology A 206 (2020) 109--124.

\bibitem{Borst-review-2023}
A.~Borst, L.~N. Groschner, How flies see motion, Annual Review of Neuroscience
  46 (2023) 17--37.

\bibitem{Rind-2024-review}
F.~C. Rind, Recent advances in insect vision in a 3d world: looming stimuli and
  escape behaviour, Current Opinion in Insect Science 63 (2024) 101180.

\bibitem{Mauss-Opposing-Motions-Fly}
A.~S. Mauss, K.~Pankova, A.~Arenz, A.~Nern, G.~M. Rubin, A.~Borst, Neural
  circuit to integrate opposing motions in the visual field, Cell 162 (2015)
  351--362.

\bibitem{LPLC-Nature-2017}
N.~C. Klapoetke, A.~Nern, M.~Y. Peek, E.~M. Rogers, P.~Breads, G.~M. Rubin,
  M.~B. Reiser, G.~M. Card, Ultra-selective looming detection from radial
  motion opponency, Nature 551 (2017) 237--241.

\bibitem{Wang-LGMD-FFI}
H.~Wang, R.~B. Dewell, Y.~Zhu, F.~Gabbiani, Feedforward inhibition conveys
  time-varying stimulus information in a collision detection circuit, Current
  Biology 28~(10) (2018) 1509--1521.
\newblock \href {https://doi.org/10.1016/j.cub.2018.04.007}
  {\path{doi:10.1016/j.cub.2018.04.007}}.

\bibitem{Zhu-LGMD-lateral-excitation}
Y.~Zhu, R.~B. Dewell, H.~Wang, F.~Gabbiani, Pre-synaptic muscarinic excitation
  enhances the discrimination of looming stimuli in a collision-detection
  neuron, Cell Reports 23~(8) (2018) 2365--2378.

\bibitem{Gabbiani2002(multiplicative-computation-LGMD)}
F.~Gabbiani, H.~G. Krapp, C.~Koch, G.~Laurent, Multiplicative computation in a
  visual neuron sensitive to looming, Nature 420~(6913) (2002) 320--324.

\bibitem{Fu-ALife-review}
Q.~Fu, H.~Wang, C.~Hu, S.~Yue, Towards computational models and applications of
  insect visual systems for motion perception: A review, Artificial Life 25~(3)
  (2019) 263--311.

\bibitem{Fu-ON/OFF-2023}
Q.~Fu, Motion perception based on on/off channels: A survey, Neural Networks
  165 (2023) 1--18.
\newblock \href {https://doi.org/https://doi.org/10.1016/j.neunet.2023.05.031}
  {\path{doi:https://doi.org/10.1016/j.neunet.2023.05.031}}.

\bibitem{Gabbiani2004(invariance-LGMD)}
F.~Gabbiani, H.~G. Krapp, N.~Hatsopoulos, C.~H. Mo, C.~Koch, G.~Laurent,
  Multiplication and stimulus invariance in a looming-sensitive neuron, Journal
  of Physiology Paris 98~(1-3 SPEC. ISS.) (2004) 19--34.

\bibitem{Contrast-polarity-mapping-2022}
R.~B. Dewell, Y.~Zhu, M.~Eisenbrandt, R.~Morse, F.~Gabbiani, Contrast
  polarity-specific mapping improves efficiency of neuronal computation for
  collision detection, eLife 11 (2022) e79772.

\bibitem{Gabbiani-2023}
F.~Gabbiani, T.~Preuss, R.~B. Dewell, Approaching object acceleration
  differentially affects the predictions of neuronal collision avoidance
  models, Biological Cybernetics 117 (2023) 129--142.

\bibitem{LGMD1-1996(Rind-neural-network)}
F.~C. Rind, D.~I. Bramwell, Neural network based on the input organization of
  an identified neuron signaling impending collision, Journal of
  Neurophysiology 75~(3) (1996) 967--985.

\bibitem{Fu-2020-Access}
Q.~Fu, H.~Wang, J.~Peng, S.~Yue, Improved collision perception neuronal system
  model with adaptive inhibition mechanism and evolutionary learning, IEEE
  Access 8 (2020) 108896--108912.

\bibitem{LGMD-car-2017(bionic-vehicle-collision)}
M.~Hartbauer, Simplified bionic solutions: A simple bio-inspired vehicle
  collision detection system, Bioinspiration {\&} Biomimetics 12~(2) (2017)
  026007.

\bibitem{jiannan-AIAI-2019}
J.~Zhao, X.~Ma, Q.~Fu, C.~Hu, S.~Yue, An {LGMD} based competitive collision
  avoidance strategy for uav, in: Artificial Intelligence Applications and
  Innovations, Springer International Publishing, 2019, pp. 80--91.

\bibitem{jiannan-TNNLS-2021}
J.~Zhao, H.~Wang, N.~Bellotto, C.~Hu, J.~Peng, S.~Yue, Enhancing lgmd's looming
  selectivity for uav with spatial-temporal distributed presynaptic
  connections, IEEE Transactions on Neural Networks and Learning Systems 34~(5)
  (2021) 2539--2553.

\bibitem{Carandini-normalization-in-neural-computation}
M.~Carandini, D.~J. Heeger, Normalization as a canonical neural computation,
  Nature Reviews neuroscience 13 (2011) 51--62.

\bibitem{Drews-dynamic-signal-compression}
M.~S. Drews, A.~Leonhardt, N.~Pirogova, F.~G. Richter, A.~Schuetzenberger,
  L.~Braun, E.~Serbe, A.~Borst, Dynamic signal compression for robust motion
  vision in flies, Current Biology 30 (2020) 209--221.

\bibitem{Fu-IJCNN-2021}
Q.~Fu, S.~Yue, Bioinspired contrast vision computation for robust motion
  estimation against natural signals, in: IEEE The International Joint
  Conference on Neural Networks, 2021, pp. 1--8.

\bibitem{Fu-Array}
Q.~Fu, Z.~Li, J.~Peng, Harmonizing motion and contrast vision for robust
  looming detection, Array 17 (2023) 100272.

\bibitem{LGMDs-2016}
F.~C. Rind, S.~Wernitznig, P.~Polt, A.~Zankel, D.~Gutl, J.~Sztarker,
  G.~Leitinger, Two identified looming detectors in the locust: Ubiquitous
  lateral connections among their inputs contribute to selective responses to
  looming objects, Scientific Reports 6 (2016) 35525.

\bibitem{Sherbakov2013}
L.~Sherbakov, A.~Yazdanbakhsh, Multiscale sampling model for motion
  integration, Journal of Vision 13~(11) (2013).

\bibitem{ZhuJ2023}
J.~Zhu, B.~Zikopoulos, A.~Yazdanbakhsh, A neural model of modified
  excitation/inhibition and feedback levels in schizophrenia, Front Psychiatry
  14 (2023) 1199690.

\bibitem{Fly2017(object-detecting-neuron)}
M.~F. Keleş, M.~A. Frye, Object-detecting neurons in drosophila, Current
  Biology 27~(5) (2017) 680--687.

\bibitem{Fly-LCs-2022}
N.~C. Klapoetke, A.~Nern, E.~M. Rogers, G.~M. Rubin, M.~B. Reiser, A
  functionally ordered visual feature map in the drosophila brain, Neuron 110
  (2022) 1700--1711.

\bibitem{DCMD-1992(selective-response-approaching)}
F.~C. Rind, P.~J. Simmons, Orthopteran {DCMD} neuron : a reevaluation of
  responses to moving objects . {I} . selective responses to approaching
  objects, Journal of Neurophysiology 68~(5) (1992) 1654--1666.

\bibitem{DCMD-1997(collision-trajectories)}
S.~Judge, F.~Rind, The locust {DCMD}, a movement-detecting neurone tightly
  tuned to collision trajectories, The Journal of Experimental Biology 200
  (1997) 2209--16.

\bibitem{LGMD1-Glayer(feature-enhancement)}
S.~Yue, F.~C. Rind, Collision detection in complex dynamic scenes using a lgmd
  based visual neural network with feature enhancement, IEEE Transactions on
  Neural Networks 17~(3) (2006) 705--716.

\bibitem{Hu-2017(Colias-LGMD1)}
C.~Hu, F.~Arvin, C.~Xiong, S.~Yue, Bio-inspired embedded vision system for
  autonomous micro-robots: The lgmd case, IEEE Transactions on Cognitive and
  Developmental Systems 9~(3) (2017) 241--254.

\bibitem{Gabbiani-2001(LGMD-invariance-angular)}
F.~Gabbiani, C.~Mo, G.~Laurent, Invariance of angular threshold computation in
  a wide-field looming-sensitive neuron, The Journal of neuroscience : the
  official journal of the Society for Neuroscience 21~(1) (2001) 314--329.

\bibitem{Gabbiani-2002(LGMD-multiplicative-computation)}
F.~Gabbiani, H.~G. Krapp, C.~Koch, G.~Laurent, Multiplicative computation by a
  looming-sensitive neuron, in: Proceedings of the second joint 24th annual
  conference and the annual fall meeting of the biomedical engineering society]
  [engineering in medicine and biology, IEEE, 2002, pp. 1968--1969.

\bibitem{Simmons-1997(LGMD2-neuron-locusts)}
P.~J. Simmons, F.~C. Rind, Responses to object approach by a wide field visual
  neurone, the {LGMD2} of the locust: Characterization and image cues, Journal
  of Comparative Physiology - A Sensory, Neural, and Behavioral Physiology
  180~(3) (1997) 203--214.

\bibitem{Qin-Fu-DNF-2024}
Z.~Qin, Q.~Fu, J.~Peng, A computational efficient and robust looming perception
  model based on dynamic neural field, Neural Networks 179 (2024).

\bibitem{Fu-2018(LGMD1-NN)}
Q.~Fu, C.~Hu, J.~Peng, S.~Yue, Shaping the collision selectivity in a looming
  sensitive neuron model with parallel {ON} and {OFF} pathways and spike
  frequency adaptation, Neural Networks 106 (2018) 127--143.
\newblock \href {https://doi.org/https://doi.org/10.1016/j.neunet.2018.04.001}
  {\path{doi:https://doi.org/10.1016/j.neunet.2018.04.001}}.

\bibitem{Fu-LGMD2-TCYB}
Q.~Fu, C.~Hu, J.~Peng, F.~C. Rind, S.~Yue, A robust collision perception visual
  neural network with specific selectivity to darker objects, IEEE Transactions
  on Cybernetics 5~(12) (2019) 5074--5088.
\newblock \href {https://doi.org/10.1109/TCYB.2019.2946090}
  {\path{doi:10.1109/TCYB.2019.2946090}}.

\bibitem{Lei-2022}
F.~Lei, Z.~Peng, M.~Liu, J.~Peng, V.~Cutsuridis, S.~Yue, A robust visual system
  for looming cue detection against translating motion, IEEE Transactions on
  Neural Networks and Learning Systems 34~(11) (2022) 8362--8376.

\bibitem{LGMD-feedback-2023}
Z.~Chang, Q.~Fu, H.~Chen, H.~Li, J.~Peng, A look into feedback neural
  computation upon collision selectivity, Neural Networks 166 (2023) 22--37.

\bibitem{Layton2015}
O.~W. Layton, A.~Yazdanbakhsh, A neural model of border-ownership from kinetic
  occlusion, Vision Research 106 (2015) 64--80.

\bibitem{Dewell-LGMD-segmentation}
R.~B. Dewell, F.~Gabbiani, Biophysics of object segmentation in a
  collision-detecting neuron, eLife~(7) (2018) e34238.
\newblock \href {https://doi.org/10.7554/eLife.34238}
  {\path{doi:10.7554/eLife.34238}}.

\bibitem{Feedforward-global-lateral-inhibition-model}
E.~G.~N. Olson, T.~K. Wiens, J.~R. Gray, A model of feedforward, global, and
  lateral inhibition in the locust visual system predicts responses to looming
  stimuli, Biological Cybernetics (2021).

\bibitem{Leveille2014}
J.~Leveille, E.~Myers, A.~Yazdanbakhsh, Object-centered reference frames in
  depth as revealed by induced motion, Journal of Vision 14~(3) (2014) 15.

\bibitem{Leveille2010}
J.~Leveille, A.~Yazdanbakhsh, Speed, more than depth, determines the strength
  of induced motion, Journal of Vision 10~(6) (2010) 10.

\bibitem{Dragonfly-2012(visual-prey-capture)}
R.~M. Olberg, Visual control of prey-capture flight in dragonflies, Current
  Opinion in Neurobiology 22~(2) (2012) 267--271.

\bibitem{hennig2008distributed}
P.~Hennig, R.~M{\"o}ller, M.~Egelhaaf, Distributed dendritic processing
  facilitates object detection: A computational analysis on the visual system
  of the fly, PLoS One 3~(8) (2008) e3092.

\bibitem{warzecha1993neural}
A.-K. Warzecha, M.~Egelhaaf, A.~Borst, Neural circuit tuning fly visual
  interneurons to motion of small objects. {I}. dissection of the circuit by
  pharmacological and photoinactivation techniques, Journal of Neurophysiology
  69~(2) (1993) 329--339.

\bibitem{geurten2007neural}
B.~R. Geurten, K.~Nordstr{\"o}m, J.~D. Sprayberry, D.~M. Bolzon, D.~C.
  O'Carroll, Neural mechanisms underlying target detection in a dragonfly
  centrifugal neuron, Journal of Experimental Biology 210~(18) (2007)
  3277--3284.

\bibitem{bolzon2009local}
D.~M. Bolzon, K.~Nordstr{\"o}m, D.~C. O'Carroll, Local and large-range
  inhibition in feature detection, Journal of Neuroscience 29~(45) (2009)
  14143--14150.

\bibitem{barnett2007retinotopic}
P.~D. Barnett, K.~Nordstr{\"o}m, D.~C. O'Carroll, Retinotopic organization of
  small-field-target-detecting neurons in the insect visual system, Current
  Biology 17~(7) (2007) 569--578.

\bibitem{Borst2011(review-motion)}
A.~Borst, T.~Euler, Seeing things in motion: Models, circuits, and mechanisms,
  Neuron 71~(6) (2011) 974--994.

\bibitem{Borst2015(common-circuit-motion)}
A.~Borst, M.~Helmstaedter, Common circuit design in fly and mammalian motion
  vision, Nature Neuroscience 18~(8) (2015) 1067--1076.

\bibitem{Maisak_2013(T4-T5-fly)}
M.~S. Maisak, J.~Haag, G.~Ammer, E.~Serbe, M.~Meier, A.~Leonhardt,
  T.~Schilling, A.~Bahl, G.~M. Rubin, A.~Nern, B.~J. Dickson, D.~F. Reiff,
  E.~Hopp, A.~Borst, A directional tuning map of drosophila elementary motion
  detectors, Nature 500~(7461) (2013) 212--216.

\bibitem{Badwan-Dynamic-nonlinearities-EMD}
B.~A. Badwan, M.~S. Creamer, J.~A. Zavatone-Veth, D.~A. Clark, Dynamic
  nonlinearities enable direction opponency in \textit{Drosophila} elementary
  motion detectors, nature neuroscience 22 (2019) 1318--1326.

\end{thebibliography}

\end{document}